\documentclass[a4paper,12pt]{article}
\PassOptionsToPackage{hypertexnames=false}{hyperref}
\usepackage{jheppub}
\usepackage{placeins}
\usepackage{graphicx,dcolumn, bm,epsfig, cancel}
\usepackage[utf8]{inputenc}
\usepackage{subfigure}
\usepackage{blindtext}
\usepackage{subfiles}
\usepackage{float}
\usepackage{amsmath}
\numberwithin{equation}{section}
\usepackage{amssymb, times}
\usepackage{subfiles}
\usepackage{color}

\title{Measuring gravitational wave spectrum from electroweak phase transition and Higgs self-couplings}
\author[a,b,c]{Shuo Guan}
\author[a,b]{Huai-Ke Guo}
\emailAdd{guohuaike@ucas.ac.cn} 
\author[d,e]{Dian Jiao}
\author[a,b]{Qingyuan Liang}
\emailAdd{qingyuan211045@gmail.com}
\author[d,e]{Lei Wu}
\emailAdd{leiwu@njnu.edu.cn}
\author[f]{Yang Zhang}
\emailAdd{zhangyang2025@htu.edu.cn}

\affiliation[a]{International Centre for Theoretical Physics Asia-Pacific (ICTP-AP), University of Chinese Academy of Sciences (UCAS), Beijing, China}
\affiliation[b]{Taiji Laboratory for Gravitational Wave Universe (Beijing/Hangzhou), University of Chinese Academy of Sciences (UCAS), Beijing, China}
\affiliation[c]{Department of Science, Science Tokyo, Tokyo 1528550, Japan}
\affiliation[d]{Department of Physics and Institute of Theoretical Physics, Nanjing Normal University, Nanjing 210023, China}
\affiliation[e]{Nanjing Key Laboratory of Particle Physics and Astrophysics, Nanjing 210023, China}
\affiliation[f]{School of Physics, Henan Normal University, Xinxiang 453007, China}
\begin{document}

\abstract
{
In this work, we demonstrate the complete process of using space-based gravitational wave detectors to measure properties of the stochastic gravitational wave background arising from a first-order electroweak phase transition. Based on frequency-domain simulations of the Taiji mission, including instrumental noise and astrophysical foregrounds, we perform parameter inference using both the Fisher information matrix and Bayesian Markov Chain Monte Carlo sampling. We show how the reconstructed spectrum constrains the macroscopic parameters of the phase transition, and further how these constraints map onto the underlying particle-physics parameters in a singlet-extended Standard Model. Our results demonstrate that the Higgs cubic and quartic self-couplings can be significantly constrained using gravitational wave observations, despite limitations arising from parameter degeneracy.
}

% \begin{document}
\maketitle

\section{Introduction}

The detection and precise measurement of the stochastic gravitational wave background (SGWB) of cosmological origin provide a unique probe of the early Universe (see, e.g., \cite{Caldwell:2022qsj,LiGong:2024qmt} for  recent reviews). Recent results from pulsar timing arrays (PTAs)~\cite{NANOGrav:2023gor,NANOGrav:2023hde,EPTA:2023sfo,Athron:2023mer} have revealed the first tantalizing evidence of a nanohertz SGWB. Ground-based interferometers, including Advanced LIGO~\cite{LIGOScientific:2017vtl}, Virgo~\cite{Kumar:2024bfe}, and KAGRA~\cite{KAGRA:2020tym}, explore the higher-frequency band from tens to thousands of hertz, and their latest observations have placed strong upper limits in the absence of a detection~\cite{LIGOScientific:2025bgj,LIGOScientific:2025kry}. Looking ahead, space-based interferometers such as LISA~\cite{LISA:2017pwj,Robson:2018ifk,LISACosmologyWorkingGroup:2022jok}, Taiji~\cite{Hu:2017mde,Ruan:2018tsw,Wu:2018clg,Zhang:2024zxj,Deng:2025syo,Ren:2023yec,Du:2025xdq,Du:2023plr,Jiang:2026lik,Wu:2025rzi}, and TianQin~\cite{TianQin:2015yph,TianQin:2020hid,Luo:2020bls} will extend the accessible frequency window into the millihertz range. Among the most intriguing cosmological sources of the SGWB are those arising from first-order phase transitions (FOPTs) in the early Universe (see, e.g.,~\cite{Caprini:2015zlo,Weir:2017wfa,Mazumdar:2018dfl,Caprini:2019egz,Bian:2021ini,Athron:2023xlk} for reviews), which offer a valuable link between cosmology and particle physics by probing symmetry breaking at different energy scales. Of particular interest is the electroweak phase transition (EWPT) associated with the breaking of the electroweak symmetry. In many extensions of the Standard Model, the EWPT can become strongly first order, generating the out-of-equilibrium conditions required for baryogenesis~\cite{Morrissey:2012db} and simultaneously producing a potentially observable gravitational wave (GW) signal within the reach of future space-based detectors~\cite{Caprini:2015zlo}.

Previous phenomenological studies of the EWPT have mostly relied on simple signal-to-noise ratio (SNR) calculations, combining the predicted SGWB with detector sensitivity curves to assess the detectability of the signal. In practice, however, experimentally extracting a faint cosmological SGWB from the observational data of a space-based detector is particularly challenging due to the presence of many noise sources and contamination from astrophysical foregrounds (see, e.g., \cite{Boileau:2020rpg,Boileau:2021sni}).  

Space-based GW detectors must contend with various sources of instrumental noise, dominated by laser frequency fluctuations and supplemented by residual acceleration noise, optical-path disturbances, etc.,~\cite{LISA:2017pwj}. In ground-based experiments, such challenges can be partly mitigated through the use of a detector network, where cross-correlation between independent data streams suppresses uncorrelated noise~\cite{Allen:1997ad,Romano:2016dpx}. Although a network of space-based detectors has been proposed~\cite{Ruan:2020smc,Cai:2023ywp,Chen:2024fto}, most current studies ---including the present work---focus on the single-detector case, which requires a different approach for signal extraction. In this context, time-delay interferometry (TDI)~\cite{Tinto:2014lxa} is employed to cancel the dominant laser frequency noise, resulting in three independent data combinations, e.g., the $A$, $E$, and $T$ channels,
each with distinct sensitivity to GWs (see, e.g., \cite{Smith:2019wny} for a recent overview). Among these, the so-called null channel $T$~\cite{Adams:2010vc} is particularly valuable: at low frequencies, it is almost insensitive to GWs, allowing it to serve as an internal noise monitor that assists in instrument calibration and in distinguishing signal from noise.

In addition, the stochastic signal of cosmological origin is blended with strong astrophysical foregrounds, most notably the unresolved galactic binaries that form a confusion background and the contribution from extragalactic compact binary coalescences (CBCs)~\cite{Tinto:2004wu,Smith:2019wny,Barack:2004wc,Caprini:2024hue,Boileau:2020rpg}. Disentangling these different components—and thereby establishing a confident detection of the SGWB from a FOPT—requires dedicated statistical analysis.
\footnote{A further challenge arises from deterministic, individually resolvable signals such as those from galactic binaries, massive black hole binaries, and extreme mass-ratio inspirals, which must be jointly analyzed along with stochastic components in the so-called global fit framework~\cite{Cornish:2005qw,Littenberg:2023xpl,Rosati:2024lcs}. This formidable task remains under active   and will not be considered in this work.}  

Significant progress has recently been achieved both in theoretical modeling and in data analysis techniques. The goal of this work is to bridge these two fronts—to integrate state-of-the-art theoretical developments with realistic experimental considerations—and to assess the achievable scientific outcomes in line with recent efforts within the community~\cite{Gowling:2021gcy,Gowling:2022pzb,Boileau:2022ter,Caprini:2024hue,Lewicki:2024xan,Huang:2025uer}.

For phenomenological studies, the interplay between GW observations and collider measurements has become a major focus of research~\cite{Ramsey-Musolf:2019lsf}. In particular, the cubic and quartic self-couplings of the Higgs boson---central to understanding the scalar potential and the mechanism of electroweak symmetry breaking, but notoriously difficult to measure at colliders~\cite{DiVita:2017vrr,Zabinski:2023jhr}---serve as key observables that highlight the complementarity between collider and GW approaches~\cite{Alves:2018jsw}.  

In this paper, we develop a framework to probe Higgs self-couplings, among others, through the space-based detection of phase-transition-induced SGWBs. Starting from a minimal benchmark Lagrangian, we compute the relevant phase transition parameters and the corresponding GW spectra over the parameter space consistent with phenomenological and theoretical constraints. We then construct a likelihood using simulated GW data in the $A$, $E$, and $T$ channels and perform parameter estimation with both the Fisher information matrix (FIM) and a Bayesian analysis based on Markov Chain Monte Carlo (MCMC) sampling, incorporating realistic astrophysical backgrounds and detector noise for a Taiji-like mission. The measured spectral uncertainties for the injected signal are subsequently mapped back to the parameter space of the particle physics model, enabling further quantitative predictions for the cubic and quartic Higgs self-couplings.  

In this work, we demonstrate the above pipeline using the simplest extension of the Standard Model that includes a real scalar singlet; the xSM~\cite{Barger:2007im, Profumo:2007wc, Profumo:2014opa, Huang:2017jws, Goncalves:2024vkj}. This study extends our previous phenomenological analyses~\cite{Alves:2018jsw} by introducing a more realistic statistical treatment tailored to space-based GW experiments.  

The remainder of this paper is organized as follows. In Sec.~\ref{dsm}, we introduce the xSM and summarize its EWPT dynamics. Section~\ref{qlc} presents the GW spectrum from the EWPT, together with the astrophysical foreground and background models considered in this work. In Sec.~\ref{djj}, we describe the detector response, noise modeling, and data analysis procedures for space-based interferometers. Section~\ref{wwh} applies the statistical inference framework to the simulated data, maps the reconstructed spectral constraints onto the xSM parameter space, and derives the corresponding implications for Higgs self-couplings. Finally, Sec.~\ref{kjh} summarizes our results and discusses their implications for future GW missions and collider experiments. In Appendix, we provide the detailed statistical inference framework, including a comparison between Fisher forecasts and Bayesian MCMC results.

\section{The model}\label{dsm}
In this section, we introduce the particle physics model and its parameter space, defined by five independent parameters. Once the model parameters are fixed, the phase transition parameters can be determined; these are the quantities that directly enter the prediction of the GW spectrum. Throughout this work, we follow the same convention as in our previous study~\cite{Alves:2018jsw}, to which we refer the reader for further details.

\subsection{xSM model}\label{ybg}

The model ``xSM'' is the simplest extension of the Standard Model and has been extensively studied in the literature due to its simplicity and its ability to accommodate a strongly FOPT~\cite{Barger:2007im, Profumo:2007wc, Profumo:2014opa, Huang:2017jws}. This model is obtained
by coupling the SM Higgs doublet with a real scalar gauge singlet, $S=v_s+s$, with the following potential:
\begin{equation}\label{potential}
\begin{aligned}
       V(H,S)=-\mu^2(H^{\dagger}H)+\lambda(H^{\dagger}H)^2+\tfrac{a_1}{2}H^{\dagger}HS \\ +\tfrac{a_2}{2}H^{\dagger}HS^2+\tfrac{b_2}{2}S^2+\tfrac{b_3}{3}S^3+\tfrac{b_4}{4}S^4,
\end{aligned}
\end{equation}
where $H$ is the SM Higgs doublet,
\begin{equation}
    H=\begin{pmatrix}
G^+
\\ 
\tfrac{1}{\sqrt{2}}(v_{\mathrm{EW}}+h+iG^0)
\end{pmatrix} ,
\end{equation}
with its vacuum expectation values $v_{\mathrm{EW}}$ and $G^+, G^0$, the Goldstone modes. Through the two minimization conditions around the vacuum ($v_{\mathrm{EW}}$,$v_s$), $\mu$ and $b_2$ are given by
\begin{equation}
\begin{split}
&\mu^2=\lambda v_{\mathrm{EW}}^2+\tfrac{1}{2}v_{s}(a_1+a_2v_{s}),
\\
& b_2=-\tfrac{1}{4v_{s}}[v_{\mathrm{EW}}^2(a_1+2a_2v_{s})+4v_{s}^2(b_3+b_4v_{s})] .
\end{split}
\end{equation}
In turn, $\lambda$, $a_1$, and $a_2$ can be replaced by the mixing angle $\theta$, the Higgs mass $m_{h_1}\backsimeq 126\ \text{GeV}$, and the mass of the heavier SM-like scalar $m_{h_2}$, from the mass matrix diagonalization,
\begin{equation}
    \begin{aligned}
        \lambda&=\tfrac{m_{h_1}^2c_{\theta}^2+m_{h_2}^2s_{\theta}^2}{2v_{\mathrm{EW}}^2},\\
        a_1&=\tfrac{2v_s}{v_{\mathrm{EW}}^2}[2v_{s}^2(2b_4+\tilde{b}_3)-m_{h_1}^2-m_{h_2}^2+c_{2\theta}(m_{h_1}^2-m_{h_2}^2)],\\
        a_2&=-\tfrac{1}{2v_{\mathrm{EW}}^2v_{s}}[-2(m_{h_1}^2+m_{h_2}^2-4b_4v_{s}^2) \\
        &\qquad + (m_{h_1}^2-m_{h_2}^2)(2c_{2\theta}v_{s}-v_{\mathrm{EW}}s_{2\theta})+4\tilde{b}_3v_s^3] ,
    \end{aligned}
\end{equation}
where $\tilde{b}_3\equiv b_3/v_s$, and $c_{\theta}\equiv\cos\theta$, $s_{\theta}\equiv\sin\theta$. Accordingly, the mass eigenstates $h_1$ and $h_2$ are linear combinations
of $h$ and $s$,
\begin{equation}
    h_1=c_{\theta}h+s_{\theta}s, \qquad h_2=-s_{\theta}h+c_{\theta}s.
\end{equation}
From the above equations, the potential in Eq.~\eqref{potential} can be fully described by the five parameters,
\begin{equation}
v_s,\qquad m_{h_2},\qquad \theta,\qquad b_3,\qquad b_4.
\label{eq:modelparas}
\end{equation}
These five parameters
% , denoted collectively as $\phi_i$ with $i=1, \cdots 5$, 
define the parameter space 
of the xSM model, which is constrained firstly by 
the theoretical requirement of the Higgs potential and by experimental measurements of various physical observables. The former includes the boundedness of the potential from below, electroweak vacuum stability at zero temperature, perturbativity, perturbative unitarity, etc. The latter includes precision measurements of the $W$ mass, the oblique electroweak corrections, the Higgs signal strengths, searches for a heavier Higgs-like scalar, etc. These have been extensively studied in the literature and our previous study~\cite{Alves:2018jsw}. Our study in this work is based on the phenomenologically viable parameter space, characterized by a large dataset of parameter space points that pass above the constraints. The readers are referred to the previous work for further details on the phenomenological constraints.

Among the physical observables that can be probed through GW observations, the Higgs self-couplings play a particularly central role. In this work, we take them as a representative example for performing statistical inference,
\begin{equation}
    \begin{split}
        i\lambda_{h_1h_1h_1}&=6\big[\lambda vc_{\theta}^3+\tfrac{1}{4}c_{\theta}^2s_{\theta}(2a_2v_s+a_1)+\tfrac{1}{2}a_2vc_{\theta}s_\theta^2 \\
        &+\tfrac{1}{3}s_\theta^3(3b_4v_s+b_3)\big],\\
        i\lambda_{h_1h_1h_1h_1}&=6(\lambda c_\theta^4+a_2s_\theta^2c_\theta^2+b_4s_\theta^4).
    \end{split}
\end{equation}
When $\theta= 0$, $i\lambda_{h_1h_1h_1}=3m_{h_1}^2/v_\mathrm{EW}$, and $i\lambda_{h_1h_1h_1h_1}=3m_{h_1}^2/v_\mathrm{EW}^2$, which are the corresponding SM values. One can also use the deviation of the Higgs self-couplings from the corresponding SM values 
$\delta\kappa_3$ and $\delta\kappa_4$,
\begin{equation}\label{hdb}
  \Delta \mathcal{L}=-\tfrac{1}{2}\tfrac{m_{h_1}^3}{v_\mathrm{EW}}(1+\delta\kappa_3)h_1^2-\tfrac{1}{8}\tfrac{m_{h_1}^2}{v_\mathrm{EW}^2}(1+\delta\kappa_4)h_1^4 .
\end{equation} 
We will later use $\delta\kappa_3$ and $\delta\kappa_4$ to assess how GW measurement can constrain Higgs self-couplings.

\subsection{Phase transition parameters}

For the GW spectrum generated by phase transitions, it is not the set of five model parameters in Eq.~\eqref{eq:modelparas} that appears directly, but rather a derived set of phase transition parameters \cite{Caprini:2024hue},
\begin{eqnarray}
T_\mathrm{n}, \quad \alpha, \quad \beta/H_\mathrm{n}, \quad v_w ,
\end{eqnarray}
which describes the thermodynamic properties of the EWPT. Specifically, $T_\mathrm{n}$ is the temperature at which the phase transition occurs, $\alpha$ quantifies the released vacuum energy normalized to the radiation energy density at $T_\mathrm{n}$, $\beta/H_\mathrm{n}$ characterizes the inverse time scale of the transition, and $v_w$ denotes the bubble wall velocity. These parameters are computed utilizing the most recent version of \texttt{PhaseTracer} with action fitting enabled~\cite{Bian:2025yfj,Athron:2024xrh,Athron:2020sbe}.

In the Bayesian statistical analysis presented later, we do not directly infer the above thermodynamic parameters. Instead, our analysis is performed on a more direct spectral model characterized by two parameters; the overall amplitude $\Omega_0$ and the peak frequency $f_{\mathrm{p}}$. This choice avoids the issue of parameter degeneracy that would arise if one were to analyze the thermodynamic parameters directly, as will be discussed in subsequent sections. Once constraints on $\Omega_0$ and $f_{\mathrm{p}}$ are obtained, we then employ the mapping procedure introduced later to translate these results into constraints on the thermodynamic parameters and further into the particle-physics model parameters defined in Eq.~\eqref{eq:modelparas}. The detailed relations between these parameter sets, which require the calculation of the finite-temperature effective potential and the associated phase transition dynamics, can be found in our previous work~\cite{Alves:2018jsw} and will not be repeated here.

Finally, we note that there remain significant theoretical uncertainties (see, e.g.,~\cite{Croon:2020cgk,Athron:2022jyi,Athron:2023rfq,Lewicki:2024xan,Zhu:2025pht}) in determining the above phase transition parameters. These arise from several sources: the construction of the finite-temperature effective potential and the nucleation rate calculation, including the issue of gauge dependence for the perturbative calculation of the effective potential \cite{Patel:2011th,Arunasalam:2021zrs,Lofgren:2021ogg,Hirvonen:2021zej}; and the calculation of the parameters $T_\mathrm{n}$, $\alpha$, $\beta/H_\mathrm{n}$, and $v_w$, where different approaches with varying levels of rigor exist~\cite{Guo:2021qcq}. For simplicity,
we do not consider these theoretical uncertainties in the following statistical analysis, and we refer to~\cite{Lewicki:2024xan} for a study of this kind based on the FIM analysis.

\section{Gravitational waves: cosmological and astrophysical} \label{qlc}

While the target is the GW signal from the EWPT, one must also account for astrophysical and other possible cosmological backgrounds or foregrounds. These additional signals contribute incoherently to the observed data, thereby increasing the challenge of detecting the target signal. As a first step and for simplicity, we include the background from extragalactic CBCs, while a more complete treatment of other additions will follow the same methodology, involving only additional computational complexity.

\subsection{Electroweak phase transition} \label{ufc}

In a FOPT, vacuum bubbles are nucleated, then grow, collide, and merge, 
driving the universe from a metastable to a stable state~\cite{Guth:1981uk,Hindmarsh:2020hop}.  
The expansion and collision of these bubbles disturb the surrounding plasma
in the form of sound waves (SWs) and magnetohydrodynamic (MHD) turbulence~\cite{Caprini:2015zlo,Hindmarsh:2013xza}.  
As a result, the SGWB from a FOPT has three main contributions;  
(i) bubble collisions~\cite{Kosowsky:1991ua,Kosowsky:1992rz,Kosowsky:1992vn,Huber:2008hg,Jinno:2016vai,Jinno:2017fby},  
(ii) SWs~\cite{Hindmarsh:2013xza,Hindmarsh:2015qta,Cai:2023guc,RoperPol:2023dzg}, and  
(iii) MHD turbulence~\cite{Kosowsky:2001xp,Caprini:2009yp,Binetruy:2012ze,Gogoberidze:2007an,Niksa:2018ofa,RoperPol:2019wvy}.  
However, the bubble-collision contribution is usually negligible compared to SWs for the EWPT, 
and that of the MHD turbulence is both subdominant and still uncertain. Thus, we
consider only the dominant contribution from SWs.

The spectrum from SWs takes the following form~\cite{Caprini:2015zlo,Guo:2020grp}:  
\begin{equation}\label{tdv}
    \begin{aligned}
        \Omega_{\mathrm{sw}}(f) h^2
        &= 2.65 \times 10^{-6}
        \left(\frac{H_{\mathrm{n}}}{\beta}\right)
        \left(\frac{\kappa_{\mathrm{sw}} \alpha}{1+\alpha}\right)^{2}
        \left(\frac{100}{g_{*}}\right)^{1/3} 
        \\&
        v_{\mathrm{w}}
        \left(\frac{f}{f_{\mathrm{p}}}\right)^{3}
        \left(\frac{7}{4+3(f/f_{\mathrm{p}})^{2}}\right)^{7/2}
        \Upsilon\!\left(\tau_{\mathrm{sw}}\right)\;,
    \end{aligned} 
\end{equation}
where $H_\mathrm{n}$ is the Hubble rate at the nucleation temperature $T_\mathrm{n}$,  
$f_{\mathrm{p}}$ is the peak frequency~\cite{Romero:2021kby},  
\begin{equation}
    f_{\mathrm{p}}
    = \frac{19}{v_{\mathrm{w}}}
    \left(\frac{\beta}{H_{\mathrm{n}}}\right)
    \left(\frac{T_\mathrm{n}}{100~\mathrm{GeV}}\right)
    \left(\frac{g_{*}}{100}\right)^{1/6}
    10^{-6}~\mathrm{Hz}\;,
\end{equation}
and $\Upsilon$ is a suppression factor~\cite{Guo:2020grp}, a function of the finite lifetime~\cite{Ellis:2018mja} of the 
SWs $\tau_{\text{sw}}$, which in a radiation-dominated universe takes the following form:  
\begin{equation}\label{tdl}
    \Upsilon(\tau_{\mathrm{sw}})
    = 1-\frac{1}{\sqrt{1+2\tau_{\mathrm{sw}} H_\mathrm{n}}} \;,
\end{equation}
while the formula for a universe with a generic expansion rate was derived recently in~\cite{Guo:2024kfk}.  

The lifetime $\tau_{\text{sw}}$ is usually chosen to be the time for the onset of MHD turbulence, $\tau_{\mathrm{sw}} = \frac{R_{\mathrm{pt}}}{\overline{U}_f}\;$~\cite{Weir:2017wfa}, 
where $R_{\mathrm{pt}}=(8\pi)^{1/3} v_{\mathrm{w}}/\beta$ is the mean bubble separation,  
and $\overline{U}_f = \sqrt{\frac{3\kappa_{\mathrm{sw}}\alpha}{4(1+\alpha)}}$ 
is the root-mean-square fluid velocity~\cite{Weir:2017wfa}.  
For convenience, one can also define $K \equiv \frac{\alpha}{1+\alpha}$~\cite{Caprini:2024hue}, 
which encapsulates the dependence on transition strength.  

From an experimental perspective, what matters most for detection are the overall amplitude and the spectral shape of the signal.
This then allows us to write the GW spectrum in the following simplified form with a minimal set of independent parameters,
\begin{equation}\label{pgv}
    \Omega_{\mathrm{sw}}(f) 
    =
    \Omega_0 
    \left(\frac{f}{f_{\mathrm{p}}}\right)^{3}
    \left(\frac{7}{4+3\left(f / f_{\mathrm{p}}\right)^{2}}\right)^{7/2} ,
\end{equation}
which is characterized by two effective parameters;  
(1) $\Omega_0$, controlling the overall amplitude, and  
(2) $f_{\mathrm{p}}$, setting the peak frequency.  
The inference of these two parameters is what directly emerges from a Bayesian analysis of observed or simulated data.  
However, this creates a challenge when translating the measured $(\Omega_0, f_{\mathrm{p}})$ back into the underlying phase transition parameters that enter the full spectrum in Eq.~\eqref{tdv} and also into the five model 
parameters in Eq.~\eqref{eq:modelparas}, which involve more parameters.  
We will return to this issue of parameter degeneracy in a later part of the analysis.

We set the spectral parameters to be log uniformly distributed in the ranges $\Omega_0 \in (10^{-20},10^{-5})$ and $f_{\mathrm{p}} \in (10^{-5},1)~\mathrm{Hz}$, as the priors, taking into account the
theoretical constraints of the xSM (as illustrated in Fig.~\ref{uem}) and 
following also~\cite{Caprini:2024hue,Romero:2021kby}.
The corresponding ranges for
phase transition parameters are $\beta/H_\mathrm{n}\in (10^{-1},10^{5})$, $T_\mathrm{n} \in (1,10^{5})~\text{GeV}$, and $K\in (10^{-3},1)$, consistent with the sampling reconstruction results shown in Fig.~\ref{fff}.
In addition, the wall velocity and the relativistic degrees of freedom are fixed to $v_w = 1.0$ and $g_* = 100$, respectively, for simplicity, and the efficiency factor $\kappa_{\text{sw}}$ is taken as a function of $\alpha$ and $v_w$ (see the appendix of~\cite{Espinosa:2010hh} for the fitting formula). We note that
choosing different values of the parameters such as $v_w$ leads to corresponding
changes in the Bayesian inference, but not qualitatively on the conclusions obtained in this work.

% These prior ranges are summarized in Table~\ref{tab:priors}.
% \begin{table}[t]
%     \centering
%     \begin{tabular}{|c|c|}
%         % \hline
%  %       \multicolumn{2}{|c|}{SWs model} \\
%         \hline
%         Parameter & Prior range (log-uniform) \\
%         \hline
%         $\beta/H_\mathrm{n}$ & $(1,10^5)$ \\
%         \hline 
%         $T_\mathrm{n}$ & $(1,10^8)\ \text{GeV}$ \\
%         \hline
%         $K$ & $(10^{-5},1)$ \\
%         \hline
%         $\Omega_0$ & $(10^{-30},10^{-5})$ \\
%         \hline
%         $f_{\mathrm{p}}$ & $(10^{-7},10)$\ \text{Hz} \\
%         \hline
%     \end{tabular}
%     \caption{\label{tab:priors}Prior values and ranges for the parameters entering the GW spectrum from SWs.}
% \end{table}

Throughout this work, we select, as a benchmark, a point in the xSM parameter space,
\begin{equation}
    v_s \simeq 34.73\ \mathrm{GeV}, \quad m_{h_2} \simeq 573.59\ \mathrm{GeV}, \quad \theta \simeq 0.099, \quad b_3 \simeq -893.83\ \mathrm{GeV}, \quad b_4 \simeq 2.39, \nonumber
\label{eq:bmxsm}
\end{equation}
which leads to the following phase transition parameters:
% \[
% \frac{\beta}{H_{n}} = 10^{2.83}, \quad T_\mathrm{n} = 10^{1.64}, \quad K = 10^{-0.404},
% \]
\[
\frac{\beta}{H_{n}} = 776.3, \quad T_\mathrm{n} = 50.16~\mathrm{GeV}, \quad K = 0.289,
\] 
and the predicted spectral parameters,
% \[
% \Omega_0 = 10^{-11.5}, \quad f_{\mathrm{p}} = 10^{-2.25}.
% \]
\[
\Omega_0 = 9.97\times10^{-13}, \quad f_{\mathrm{p}} = 7.4\times10^{-3}~\mathrm{Hz}.
\]
We will simulate the SGWB with this benchmark point, and then investigate how well it can be reconstructed with the 
simulated data of the detector in the presence of astrophysical background and detector noises, which will be discussed in the next section.

\subsection{Astrophysical foreground and background}

CBCs, both within our Galaxy and beyond, produce numerous signals with low SNRs that overlap incoherently. Their superposition gives rise to two main components in the frequency range of space-based GW detectors; a foreground from unresolved galactic binaries, and a background from extragalactic CBCs \cite{Boileau:2021sni}.

In the simplified case, the contribution from the astrophysical background can be captured by a power-law model~\cite{Boileau:2020rpg},
\begin{equation} \label{hbf}
    \Omega_{\text{GW,ast}}(f) = \Omega_{\text{ast}}\left(\frac{f}{f_{\text{ref}}}\right)^\varepsilon ,
\end{equation}
where $f_{\text{ref}}=1\ \text{mHz}$ is the reference frequency, and $\varepsilon$ is the spectral index.  
For a background generated by the incoherent superposition of many binary black hole and neutron star mergers, one typically expects 
that $\varepsilon = \tfrac{2}{3}$.  
In our simulations, we inject such a background with an amplitude $\Omega_{\text{ast}} \simeq 3.16\times10^{-12}$, chosen within the log-uniform prior range $(10^{-14}, 10^{-10})$, consistent with the parameter space considered in Refs.~\cite{LIGOScientific:2019vic, Boileau:2022ter}.

On the other hand, the foreground is mainly due to unresolved galactic white dwarf binaries, which dominate the low-frequency band and form a confusion signal~\cite{Barack:2004wc,Adams:2013qma,Liu:2023qap}. For simplicity, in this work we include only the extragalactic astrophysical background and neglect the galactic foreground.

\section{Detection of SGWB with space-based detectors}\label{djj}

Here we present the method used for simulating and inferring the SGWB. Space-based GW detectors---such as Taiji, LISA, and TianQin---are designed as triangular constellations of three spacecraft connected by laser links~\cite{Gowling:2022pzb}, with each arm length corresponding to the distance between two free-falling test masses~\cite{Gowling:2021gcy}. A passing GW induces tiny differential changes in these arm lengths, which modify the interference pattern recorded by the spacecraft.  

The detection of SGWB can be performed either with a single detector or a network of detectors~\cite{Ruan:2020smc,Wang:2021njt}. In this work, we focus on the single-detector case, which differs conceptually from the standard method used by ground-based detectors such as LIGO/Virgo/KAGRA~\cite{KAGRA:2021kbb}, where cross-correlation between multiple detectors is employed to suppress uncorrelated noises~\cite{Allen:1997ad,Romano:2016dpx}. For a space-based constellation, the noises in the different data channels are correlated and cross-correlation is not applicable; instead, a null channel~\cite{Adams:2010vc} can be constructed that is insensitive to GWs and thus provides a probe of instrumental noise, while the remaining channels retain GW sensitivity and can be used to search for the SGWB. We adopt this null-channel method in our analysis.  

To evaluate the detection prospects for SGWB from a phase transition, we inject the simulated SGWB signal corresponding to the benchmark in Eq.~\eqref{eq:bmxsm} into the simulated detector noise data and attempt to infer both the spectral shape as defined by Eq.~\eqref{pgv} and the corresponding model parameters. In what follows, we first introduce the detector response and main noise sources, then describe the simulation of signal and noise, and finally construct the likelihood function used for statistical inference in the next section.

\subsection{Signal response and noise models}

Here we provide a brief introduction to the signal response and noise models for space-based GW detectors, and refer readers to Ref.~\cite{Smith:2019wny} for more detailed discussions of this topic. The GWs are characterized by their amplitude $h_{ij}$, or more appropriately by the spectral density $\Omega_{\text{GW}}(f)$ for SGWB. The detector measures the signals as three streams of data $d_{i}(t)$, where $i = 1, 2, 3$ represent $A_{BC}$, $B_{CA}$, and $C_{AB}$ respectively, and $A, B, C$ label the spacecraft forming the triangular interferometric constellation. These data streams capture the relative optical path change of a Michelson interferometer consisting of two adjacent arms. Due to the properties of the SGWB, the signal amplitudes $d_{1/2/3}(t)$ are random variables. We will work mainly in the frequency domain, with the following convention for the Fourier transform:
\begin{eqnarray}
\label{tfv}
d(t)&=&\int_{-\infty}^{\infty} \mathrm{d} f ~ \tilde{d}(f) \exp (2 \pi \mathrm{i} f t), 
\nonumber \\
\tilde{d}(f)&=&\int_{-\infty}^{\infty} \mathrm{d} t ~ d(t) \exp (-2 \pi \mathrm{i} f t). 
\end{eqnarray}
Given these three data streams in the frequency domain, a diagonalization of the covariance matrix can be performed such that three orthogonal channels, commonly denoted as $A$, $E$, and $T$, can be obtained with the following correlators:
\begin{equation}\label{ugc}
\langle \tilde{d}_I(f) \tilde{d}^{\ast}_J(f^{\prime})\rangle = \frac{1}{2} P_{IJ}(f) \delta_{IJ} \delta(f - f^{\prime}) ,
\end{equation}
where $I,J$ denotes the $A$, $E$, and $T$ channels, and these channels are related to the original ones by the following relations:
\begin{equation}
    \left\{
        \begin{aligned}
            \tilde{d}_A&=\frac{1}{\sqrt{2}}\left(\tilde{d}_3-\tilde{d}_1\right) ,\\
            \tilde{d}_E&=\frac{1}{\sqrt{6}}\left(\tilde{d}_1-2\tilde{d}_2+\tilde{d}_3\right), \\
            \tilde{d}_T&=\frac{1}{\sqrt{3}}\left(\tilde{d}_1+\tilde{d}_2+\tilde{d}_3\right) .
        \end{aligned}
    \right.
\end{equation} 
The one-sided power spectral density (PSD) $P_{IJ}(f)$ is now a diagonal matrix, allowing the simplified notation $P_a$ $(a = A, E, T)$. $P_a$ receives contributions from both the instrumental noise and the GW signal, 
\begin{equation}\label{udd}
    P_{a}(f) = S_a(f) + N_a(f) .
\end{equation}

The signal contribution in channel $a$ is related to the GW energy density spectrum $\Omega_{\mathrm{GW}}(f)$ through
\begin{equation}\label{esc}
    S_a(f) =
    \frac{3 H_0^2}{4 \pi^2}
    \frac{\Omega_{\mathrm{GW}}(f)}{f^3}
    \mathcal{R}_a(f),
\end{equation}
where $\mathcal{R}_a(f)$ is the detector response function.

In the idealized equal-arm configuration adopted in this work, the analytical expressions for the response functions read~\cite{Smith:2019wny,Cornish:2001bb}
\begin{equation}
    \begin{aligned}
        \mathcal{R}_{A}(f) &= \mathcal{R}_{E}(f) =
        \frac{9}{5} |W(f)|^2
        \left[1 + \left(\frac{f}{4 f_{\ast}/3}\right)^2\right]^{-1}, \\
        \mathcal{R}_{T}(f) &=
        \frac{1}{1008} |W(f)|^2
        \left(\frac{f}{f_{\ast}}\right)^6
        \left[1 + \frac{5}{16128}
        \left(\frac{f}{f_{\ast}}\right)^8\right]^{-1}.
    \end{aligned}
\end{equation}

Here, we adopt the convention $\Delta L / L$, where $\Delta L$ is the difference between the optical path lengths of the two interferometer arms and $L$ is the single-arm length. This differs from the convention $\Delta L / L_\delta$ used in Ref.~\cite{Smith:2019wny}, where $L_\delta$ denotes the round-trip light travel distance of the two paths in the Michelson configuration. As a result, our response functions differ by an overall factor of $1/4$ compared to those in Ref.~\cite{Smith:2019wny}. Both conventions are physically equivalent, provided that the same normalization is used consistently for the response functions and the noise PSD.

\begin{figure}[t]
\centering
\includegraphics[width=0.45\textwidth]{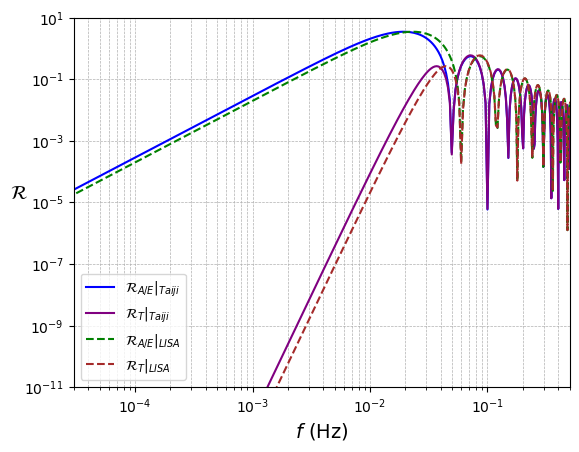}
\quad
\includegraphics[width=0.47\textwidth]{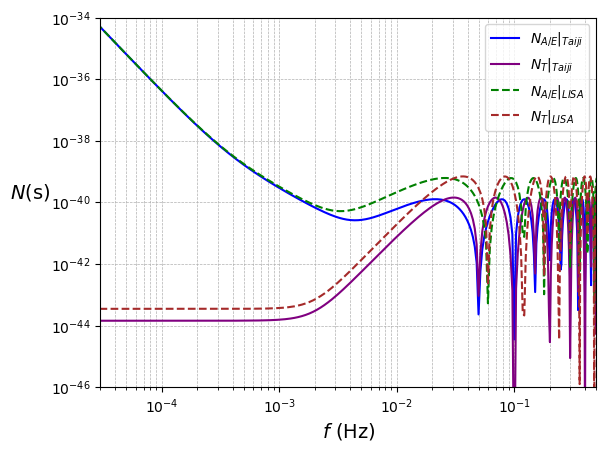}
\caption{
Comparison of the response functions (left) for the $A$, $E$, and $T$ channels of Taiji (solid curves) and LISA (dashed curves), together with their respective noise spectral densities (right). 
}
\label{urh}
\end{figure}

The primary source of detector noise arises from fluctuations in the laser frequency combined with unequal interferometer arm lengths. This noise component is suppressed by approximately eight orders of magnitude through the use of TDI~\cite{Tinto:2004wu}, which constructs appropriate linear combinations of raw measurements to achieve laser noise cancelation. For simplicity, we assume equal and fixed arm lengths in this analysis. The application of TDI introduces the factor $W = 1 - e^{-2 i f/f_\ast}$, where $f_{\ast} = \tfrac{c}{2 \pi L}$, $c$ denotes the speed of light, and $L$ represents the arm length of the detectors, which is $3 \times 10^9\ \text{m}$ for Taiji and $2.5 \times 10^9\ \text{m}$ for LISA. The response functions for the Taiji and LISA detectors are presented in Fig.~\ref{urh}. At low frequencies, the $T$ channel exhibits a significantly weaker response to GW signals compared to the $A$ and $E$ channels, and therefore serves as an effective null channel for GW detection.

%%%%%%%%%%%%%%%%%%%%%%%%%%%%%%%%%%%%%%%%%%%%%%%%%%%%%%%%%%%%%%%%%%
%%%%%%%%%%%%%%%%%%%%%%%%%%%%%%%%%%%%%%%%%%%%%%%%%%%%%%%%%%%%%%%%%%

With the signal contribution now determined, we need the PSD of each of the major noise components $N_a(f)$. For both Taiji and LISA, the noise models share a similar form (see e.g., \cite{Babak:2021mhe}) and consist of two primary components. One is the acceleration noise, given by
\begin{equation}
    \sqrt{S_{\text{acc}}(f)} = N_{\text{acc}} 
     \sqrt{1 + \left(\frac{0.4\ \text{mHz}}{f}\right)^2} \sqrt{1 + \left(\frac{f}{8\ \text{mHz}}\right)^4} 
     \left(\frac{\text{m}}{\text{s}^2 \sqrt{\text{Hz}}}\right),
\end{equation}    
which is identical for both detectors, with $N_{\text{acc}} = 3 \times 10^{-15}$. The other is the optical metrology noise, expressed as
\begin{equation}
\sqrt{S_{\text{OMS}}(f)} = \delta x 
\sqrt{1 + \left(\frac{2\ \text{mHz}}{f}\right)^4}
\left(\frac{\text{m}}{\sqrt{\text{Hz}}}\right),
\end{equation}
where $\delta x = 8 \times 10^{-12}$ for Taiji and $\delta x = 15 \times 10^{-12}$ for LISA. These two components contribute to the noise PSD of the $A$, $E$, and $T$ channels,
\begin{equation}\label{qrf}
    \begin{aligned}
        N_{A} &= N_E = N_1 - N_2, \\ 
        N_T &= N_1 + 2 N_2, 
    \end{aligned}
\end{equation} 
where $N_1(f)$ and $N_2(f)$ describe the noise contributions before diagonalization, given by
\begin{equation}
    \begin{aligned}
        N_1(f) &= \frac{1}{L^2}  \left\{4 S_{\text{OMS}}(f) + 8 \left[1 + \cos^2\left(\frac{f}{f_{\ast}}\right)\right] \frac{1}{(2 \pi f)^4}S_{\text{acc}}(f) \right\} |W(f)|^2, \\
        N_2(f) &= - \frac{1}{L^2} \left[2 S_{\text{OMS}}(f) + \frac{8}{(2 \pi f)^4} S_{\text{acc}}(f)\right]\cos\left(\frac{f}{f_{\ast}}\right) |W(f)|^2 .
    \end{aligned}
\end{equation}
We show $N_{A/E/T}$ for Taiji and LISA in the right panel of Fig.~\ref{urh}. In the following simulation, we fix the noise model parameters to be that of Taiji, while the case for LISA will lead to similar yet slightly different results.

\subsection{Data simulation and signal injection}\label{fen}

Data-cleaning and preprocessing steps are required during the actual operation of space-based detectors, such as removing data gaps from interruptions in detector operation, which result in a set of time segments $N_0$, each with a duration $T$. These segments collectively span a total observation time $T_t = N_0 T$. Here, we choose simply $T = 10^6~\mathrm{s}$ (approximately 11.4 days) following~\cite{Caprini:2024hue} and $N_0 = 126$, so that the total effective observing duration is about four years. This segmentation approach is also convenient for frequency-domain analyzes since each segment can be assumed to be approximately stationary. 

To accurately reconstruct signals and avoid distortion, the sampling frequency $f_s$ must satisfy $f_s > 2 f_{\text{max}}$, where $f_{\text{max}}$ is the maximum frequency of interest, as dictated by the Nyquist sampling theorem. The data are represented as an equally spaced time series with a time interval of $\Delta t$ between samples, resulting in a sampling rate of $f_s = \tfrac{1}{\Delta t}$.
In Fourier transformation analysis, the data from a single observation period are integrated, either over the range $0$ to $T$ or from $-\tfrac{T}{2}$ to $\tfrac{T}{2}$, to transform into the frequency domain. The minimal frequency resolution achievable in this domain is $\Delta f = \tfrac{1}{T}$, and the maximum analyzable frequency is then $f_{\text{max}} = \tfrac{1}{2 \Delta t}$, where $N = \tfrac{T}{\Delta t}$ is the total number of data points within the observation period. In this research, we select $\Delta t = 1\ \text{s}$ for convenience, and the frequency range is $\left[3 \times 10^{-5}, 0.5\right]\ \text{Hz}$.

In practical applications, the data obtained in the time domain are not continuous, so the discrete Fourier transform is used,
\begin{equation}\label{trd}
    \left\{
        \begin{aligned}
            \tilde{d}(f_k)&=\sum\limits_{n=1}^{N}d(t_\mathrm{n}) e^{-i2 \pi f_k t_\mathrm{n}} , \\
            d(t_\mathrm{n})&=\frac{1}{N}\sum\limits_{k=1}^{N}\tilde{d}(f_k) e^{2 \pi i f_k t_\mathrm{n}} .
        \end{aligned} 
    \right.
\end{equation} 
where $f_k = \tfrac{k}{N \Delta t}$ and $t_\mathrm{n} = n \Delta t$ represent the discrete frequency and time points, respectively. With this convention, the relation between $\tilde{d}(f)$ and $\tilde{d}(f_k)$ becomes $\tilde{d}(f) = \tilde{d}(f_k)\, \Delta t$. Thus, Eq.~\eqref{ugc} can be reformulated as \cite{Pieroni:2020rob}
\begin{equation}\label{yrc}
        \left\langle \tilde{d_I}(f_{k}) \tilde{d}^{*}_J(f_{k'})\right\rangle 
        =
        \frac{T f_s^2}{2}P_{IJ}\left(f_{k}\right) \delta_{kk'} .
\end{equation}
Given that $\mathbf{P}$ is a diagonal matrix and $\tilde{d}(f)$ is uncorrelated at different frequency points, the expression above can be simplified as
\begin{equation}\label{yhc}
        \left\langle \tilde{d_a}(f_{k}) \tilde{d}^{*}_a(f_{k})\right\rangle 
        =
        \frac{Tf_s^2}{2}P_{a}\left(f_{k}\right) ,
        \,
        (a=A,E,T).
\end{equation}
%%%%%%%%%%%%%%%%%%%%%%%%%%%%%%%%%%%%%%%%%%%%%%%%%%%%%%%%%%%%%%%%%%%%%%%
\begin{figure}[t]
\centering
\includegraphics[width=0.45\columnwidth]{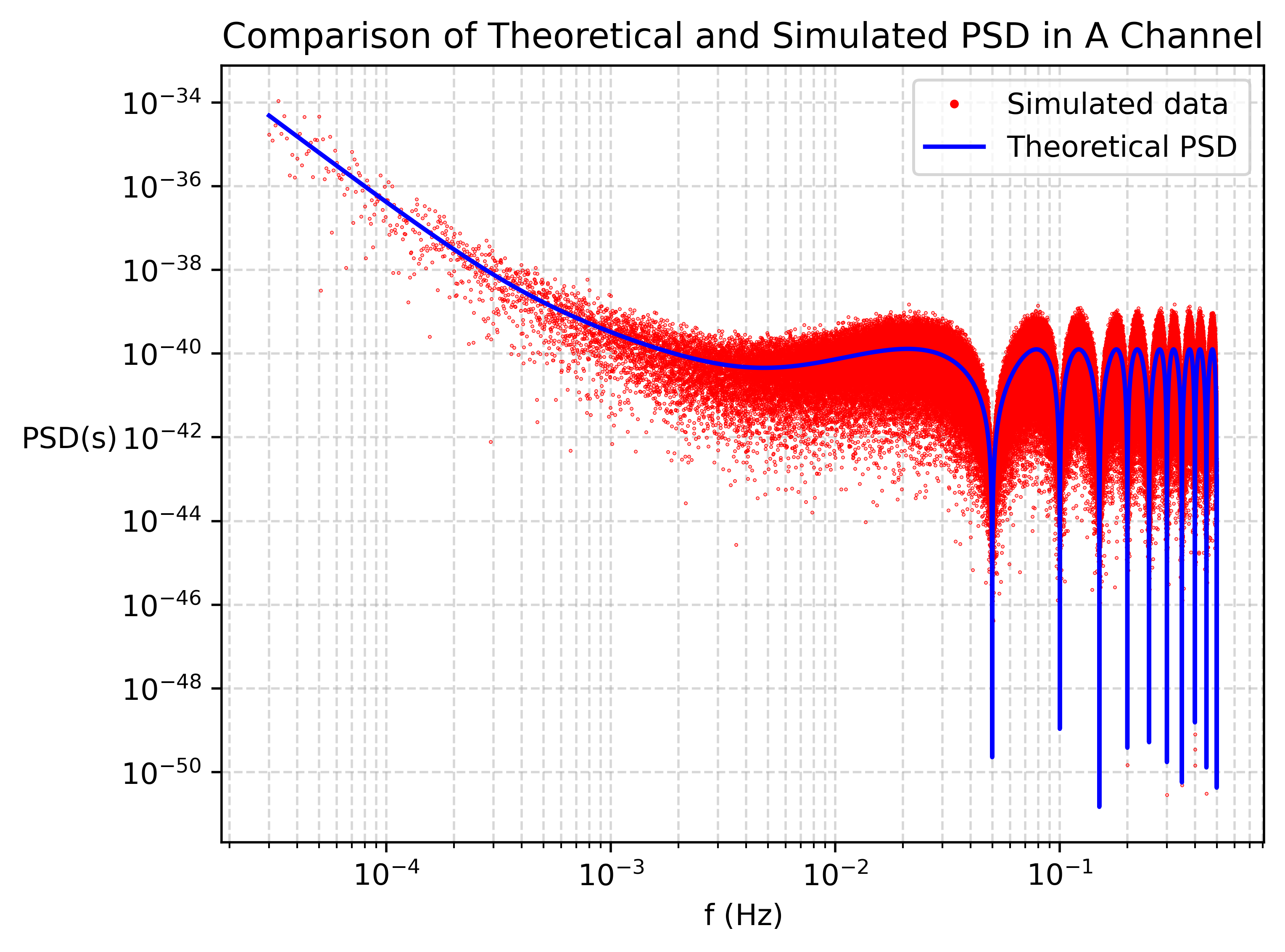}
\quad
\includegraphics[width=0.45\columnwidth]{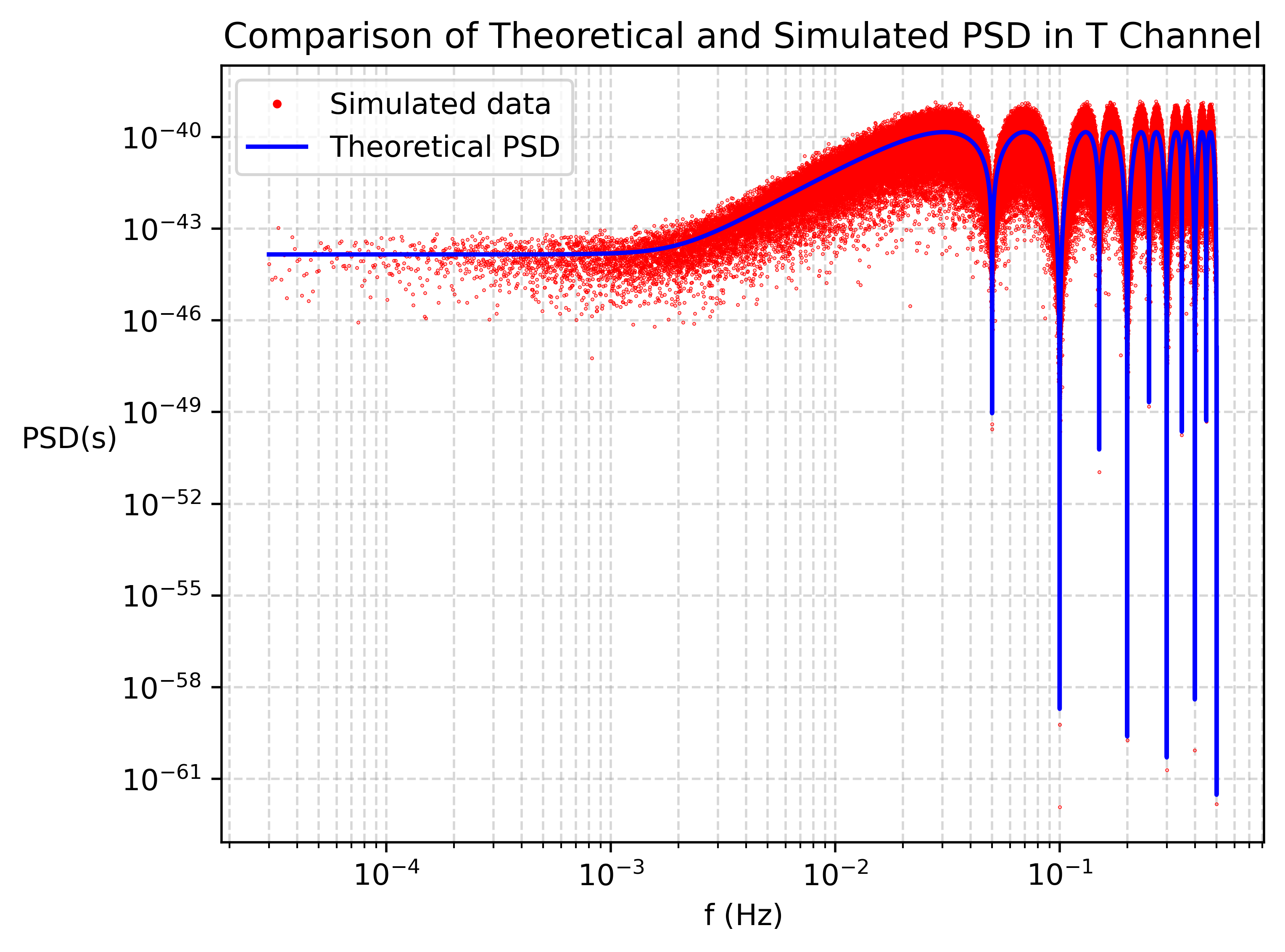}
\caption{
This figure shows the theoretical PSD curve (blue) and the simulated data in channels $A$ and $T$ (red points). The curve is computed using Eqs.~\eqref{udd} and \eqref{esc}, while the red points are generated from Eq.~\eqref{hdm}. The analysis focuses on the frequency range $\left[3 \times 10^{-5}\,\text{Hz},\,0.5\,\text{Hz}\right]$, with $\Omega_\text{ast} = 10^{-11.5}$ and $\varepsilon = 2/3$.
} 
\label{ygd}
\end{figure} 
%%%%%%%%%%%%%%%%%%%%%%%%%%%%%%%%%%%%%%%%%%%%%%%%%%%%%%%%%%%%%%%%%%%%%%%
For $\tilde{d}(f)$, which is a complex quantity, its real and imaginary parts are statistically independent and follow the same Gaussian distribution \cite{Maggiore:2018sht}, with zero mean and variances given by
\begin{equation}\label{oqb}
    \begin{split}
        \left\langle \left( \text{Re}~\tilde{d}_a(f_k)\right)^2 \right\rangle
        &=
        \left\langle \left( \text{Im}~\tilde{d}_a(f_k)\right)^2 \right\rangle
        \\
        &=
        \frac{Tf_s^2}{4}P_a(f_k) \equiv 
        \sigma_a^2(f_k) .
    \end{split}
\end{equation}
More explicitly, the real part follows the Gaussian distribution:
\begin{equation}\label{scs}
    P\left( \text{Re}~\tilde{d}_a(f_k)\right) 
    =
    \frac{1}{\sqrt{2\pi \sigma_a^2}} \exp\left[{-\frac{\left( \text{Re}~\tilde{d}_a(f_k)\right)^2}{2 \sigma_a^2}}\right]  ,
\end{equation}
and similarly for the imaginary part. Therefore, the probability of observing the complex value $\tilde{d}_a(f_k)$ is
\begin{equation}\label{hdm} 
    P\left(\tilde{d}_a(f_k)\right) 
    =
    \frac{1}{2\pi \sigma_a^2} 
    \exp\left[{-\frac{\left|\tilde{d}_a(f_k)\right|^2}{2 \sigma_a^2}}\right] .
\end{equation} 
With this understanding, we can now simulate the observational data in the $A$, $E$, and $T$ channels, including the above-mentioned instrumental noise and an injection of SGWB. In Fig.~\ref{ygd}, we show the PSD in blue for the $A$ and $T$ channels in the left and right panels, respectively, and display one realization of the simulated data as red dots. Note that when averaging over different realizations of the simulated data is performed, the red dots converge to the theoretical blue curve, implying the correctness of the simulated data.

\subsection{Likelihood}
The amount of data in the time domain is $N$; however, according to Fourier transformation theory, only $\tfrac{N}{2}$ (complex) data points are independent in the frequency domain. Thus, the likelihood for the $A$ channel can be expressed as
\begin{equation}
    \mathcal{L}
    =
    \prod_{k=1}^{N/2}
    \frac1{2\pi \sigma_a^2}
    \exp\left[{-\frac{\left|\tilde{d}_a(f_k)\right|^2}{2 \sigma_a^2}}\right] .
\end{equation}   
Combining the three channels, the likelihood for a single segment is given by 
\begin{equation}\label{raw}
    \mathcal{L}
    =
    \prod_{k=1}^{N/2}\frac1{8\pi^{3}\sigma_A^2\sigma_E^2\sigma_T^2}
    \exp
    \left[
    - \sum_{a=A,E,T}\frac{\left|\tilde{d}_a(f_k)\right|^2}{2 \sigma_a^2}
    \right].
\end{equation}
Considering all the segments requires a separate label to represent the segment index, i.e., $\tilde{d}_a^\kappa(f_k)$ with $\kappa = 1, 2, 3, \cdots, N_0$, and the likelihood of observing data from all segments becomes
\begin{equation}\label{gdh}
    \mathcal{L}
    =
    \prod_{\kappa=1}^{N_0} \prod_{k=1}^{N/2}
    \frac1{8\pi^{3}\sigma_A^2\sigma_E^2\sigma_T^2}
    \exp
    \left[
    - \sum_{a=A,E,T}\frac{\left|\tilde{d}_a^\kappa(f_k)\right|^2}{2 \sigma_a^2}
    \right].
\end{equation}
Its logarithmic form can be expressed as
\begin{equation}\label{sjc}
    \begin{split}
        \ln\mathcal{L}
        =-\sum_{\kappa=1}^{N_0}\sum\limits_{k=1}^{N/2} 
        \Bigg\{
            &\ln\frac{\pi^3 T^3 f_s^6\left[S_A(f_k) + N_A(f_k)\right]       \left[S_E(f_k) + N_E(f_k)\right] N_T(f_k)}{8} 
            \\
            &+
            \frac{2}{Tf_s^2}
            \left[
                \frac{\left|\tilde{d}^\kappa_A(f_k)\right|^2}{S_A(f_k)+N_A(f_k)}
                +
                \frac{\left|\tilde{d}^\kappa_E(f_k)\right|^2}{S_E(f_k)+N_E(f_k)}
                +
                \frac{\left|\tilde{d}^\kappa_T(f_k)\right|^2}{N_T(f_k)}
            \right]
        \Bigg\} .        
    \end{split}
\end{equation} 
In practical analysis, the amount of data may be too large to handle directly, which necessitates the use of approximation methods. Therefore, in the literature, two approximation methods are usually considered based on the central limit theorem and coarse graining~\cite{Caprini:2024hue}. In this work, we use the full likelihood above directly.

\subsection{Statistical inference framework}\label{mbl}

To extract physical information from the simulated SGWB data, we employ two complementary statistical approaches; the FIM and Bayesian inference based on MCMC sampling.

The FIM provides a fast and computationally efficient method to estimate parameter uncertainties by approximating the likelihood function as a multivariate Gaussian in the vicinity of its maximum. Under this assumption, the inverse of the FIM yields the Cramér-Rao lower bound (CRLB), which represents the minimum achievable variance for any unbiased estimator. This method is particularly useful for forecasting sensitivities, estimating relative uncertainties, and visualizing parameter correlations through confidence ellipses.

In contrast, Bayesian inference does not rely on Gaussian or linear assumptions. Instead, it samples the full posterior probability distribution of the model parameters, allowing for a faithful characterization of non-Gaussian features, parameter degeneracies, and correlations. While computationally more demanding, MCMC sampling provides robust and statistically complete uncertainty estimates.

In this work, the FIM is primarily used for fast forecasts and qualitative understanding, while MCMC sampling is adopted to obtain the final and reliable parameter constraints. The detailed mathematical formulation and derivations underlying these statistical methods are presented in Appendix~\ref{app:fim}.

\section{Results}\label{wwh}

With the statistical analysis framework established in the preceding sections, we are now ready to investigate, via simulation, how the parameters in our model can be measured. This study is conducted at multiple levels of complexity. We begin with a simulation that includes only detector noise and a SGWB of astrophysical origin. Subsequently, we inject a SGWB from the SWs, as introduced in Sec.~\ref{qlc}. The goal is to infer the model parameters and quantify the associated uncertainties. The inferred results are then mapped onto the xSM parameter space and used to extract physical observables, such as the Higgs self-couplings.

\subsection{Measurements of the astrophysical parameters} 

We start with a simpler scenario in which only instrumental noises and the astrophysical background are present in the simulated data, and we estimate the corresponding parameters of this model, namely the two noise parameters, along with the amplitude and spectral index of the astrophysical background in Eq.~\eqref{hbf}. Inserting the spectrum of the astrophysical background into Eq.~\eqref{esc}, we obtain the following PSD:
\begin{equation}
    S_{A,E}(f) = \frac{3 H_{0}^{2}}{4 \pi^{2}} \frac{\Omega_{\text{ast}} \left( \tfrac{f}{f_{\text{ref}}} \right)^{\varepsilon}}{f^{3}} \, \mathcal{R}_{A,E}(f) \,,
\end{equation}
where $\Omega_{\text{ast}}$ denotes the amplitude of the SGWB spectrum, and the reference frequency is chosen as $f_{\text{ref}} = 1\,\text{mHz}$. The function $\mathcal{R}_{A,E}(f)$ represents the detector response function in the $A$ or $E$ channel. The corresponding likelihood function is given by Eq.~\eqref{sjc}.

Firstly, the FIM is employed to analyze the parameters in the spectrum, commonly denoted as ${\bf \theta}$ and to derive the corresponding confidence ellipses. By substituting Eq.~\eqref{sjc} into Eq.~\eqref{wmd}, the FIM can be written in the following form \cite{Boileau:2020rpg}:
\begin{equation}\label{reu}
    \begin{split}
        F_{ij}^\text{likelihood}=&N_0\sum_{k=1}^{N/2}
        \left[
        \frac{2}{[S_A(f_k)+N_A(f_k)]^2}\frac{\partial[S_A(f_k)+N_A(f_k)]}{\partial\theta_i}\frac{\partial[S_A(f_k)+N_A(f_k)]}{\partial\theta_j}
        \right.
        \\&
        \left.
        +\frac{1}{N_T^2(f_k)}\frac{\partial N_T(f_k)}{\partial\theta_i}\frac{\partial N_T(f_k)}{\partial\theta_j}
        \right] .
    \end{split} 
\end{equation} 

We then compare the confidence regions obtained via the FIM formalism with those derived from full Bayesian inference using MCMC sampling. This comparison serves as a critical validation step for the Gaussian approximation inherent in the FIM approach and highlights its possible limitations when applied to realistic GW data analysis. As discussed earlier, the FIM provides an efficient means of forecasting parameter uncertainties, assuming that the posterior distribution is well-approximated by a multivariate Gaussian centered on the maximum-likelihood estimate. 

\begin{figure}[t]
    \centering
    \includegraphics[width=0.8\textwidth]{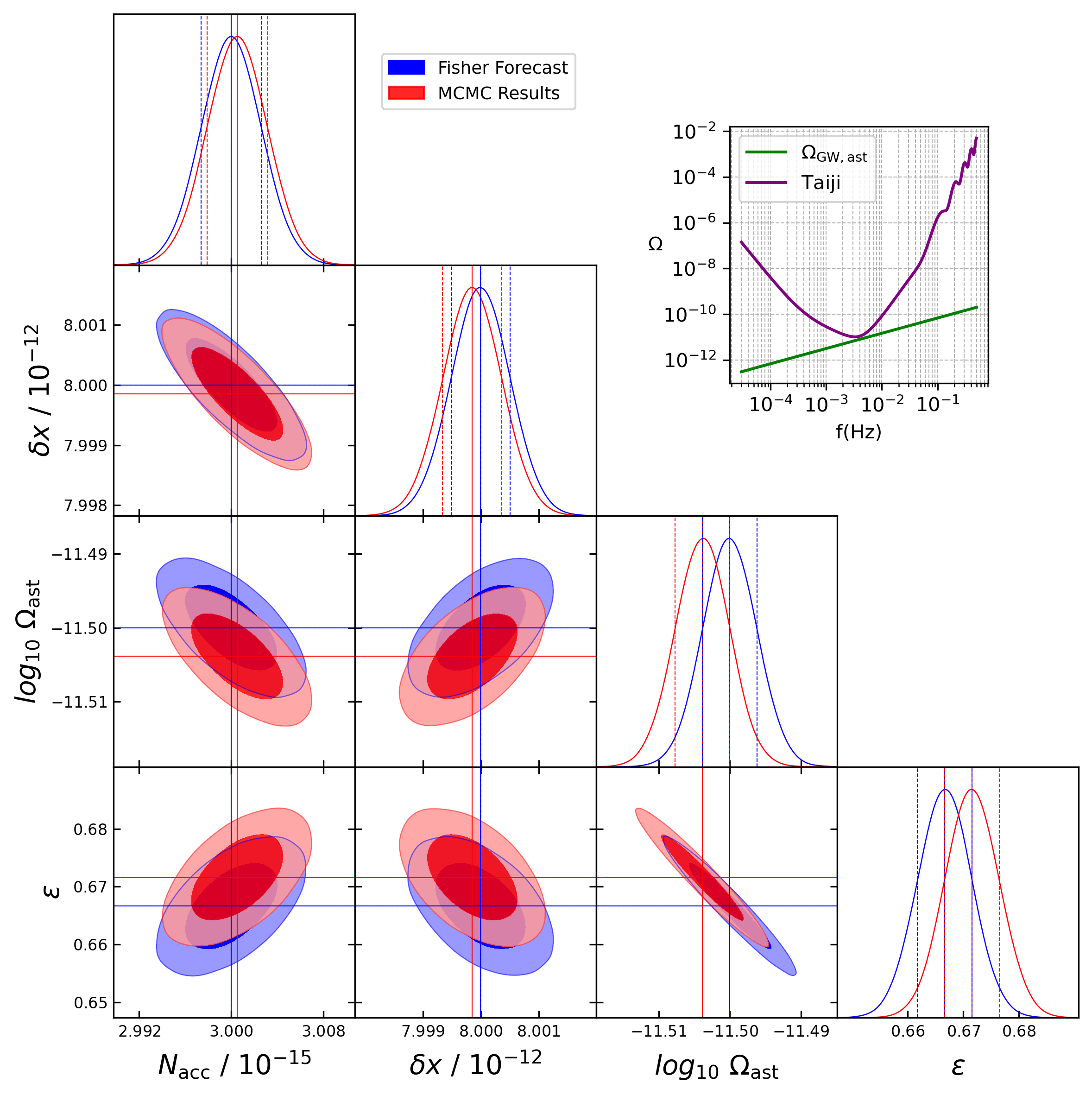}
    \caption{
    Comparison of parameter estimation uncertainties obtained from the FIM (blue) and MCMC sampling (red) for four representative parameters: two instrumental noise parameters, $N_{\mathrm{acc}}$ and $\delta x$, and two astrophysical parameters, $\Omega_{\mathrm{ast}}$ and $\varepsilon$. The injected values, $N_{\mathrm{acc}} = 3 \times 10^{-15}$, $\delta x = 8 \times 10^{-12}$, $\log_{10} \Omega_{\mathrm{ast}} = -11.5$, and $\varepsilon = 2/3$, are indicated by solid blue lines, while the MCMC-recovered best-fit values are shown as solid red lines. The dark and light-blue shaded regions denote the 68\% and 95\% confidence contours predicted by the FIM, respectively, whereas the red contours represent the corresponding credible regions from MCMC sampling. The diagonal panels display the marginalized one-dimensional posterior distributions from both approaches, with dashed vertical lines marking the $1\sigma$ intervals (blue for FIM and red for MCMC). The relative uncertainties predicted by the FIM are approximately 0.088\%, 0.006\%, 0.033\%, and 0.738\%, while those obtained from the MCMC posteriors are nearly identical: 0.088\%, 0.006\%, 0.033\%, and 0.728\%. The shape and orientation of the ellipses illustrate the correlations between parameters; elongated and tilted contours indicate strong degeneracies, whereas more circular contours suggest weaker coupling.
    }
    \label{yfd}  
\end{figure}

\begin{figure}[t]
    \centering
    \includegraphics[width=0.8\textwidth]{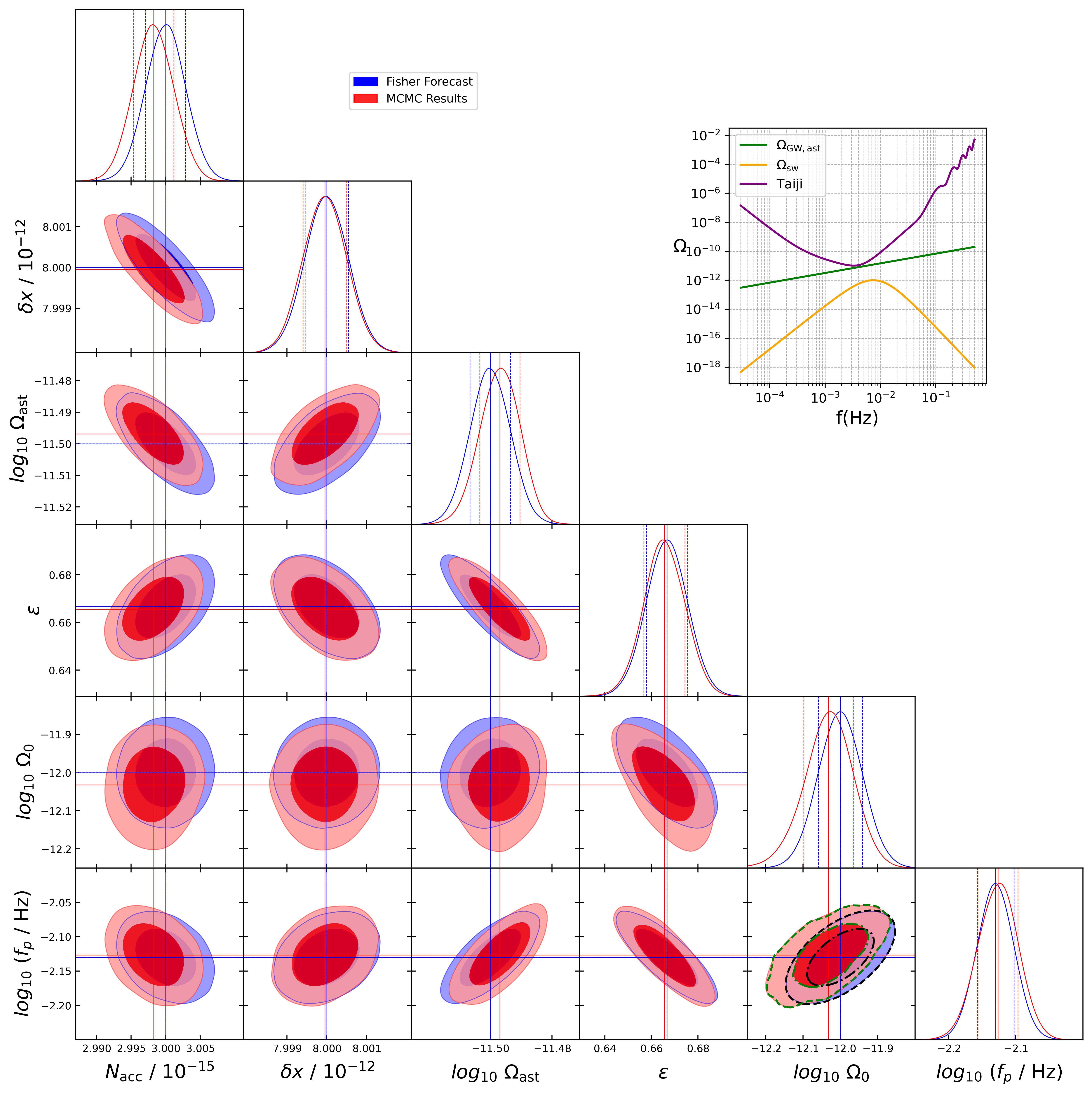}
    \caption{
    Joint confidence contours for six model parameters: $N_{\text{acc}}$, $\delta x$, $\Omega_{\text{ast}}$, $\varepsilon$, $\Omega_0$, and $f_{\mathrm{p}}$, representing instrumental noise, astrophysical background, and phase-transition contributions, respectively. The injected values, used consistently for both data generation and Fisher analysis, are $N_{\text{acc}} = 3 \times 10^{-15}$, $\delta x = 8 \times 10^{-12}$, $\log_{10} \Omega_{\rm ast} = -11.5$, $\varepsilon = 2/3$, $\log_{10} \Omega_0 = -12.0$, and $\log_{10} (f_p / \rm Hz) = -2.13$. Forecasted relative uncertainties from the FIM are 0.096\%, 0.007\%, 0.057\%, 1.333\%, 0.491\%, and 1.29\%, while those inferred from MCMC are 0.097\%, 0.007\%, 0.057\%, 1.329\%, 0.552\%, and 1.406\%. Blue contours denote the 68\% and 95\% confidence regions from the FIM, while red contours show the corresponding MCMC credible regions. Vertical blue lines indicate the injected (true) parameter values, whereas vertical red lines mark the MCMC-recovered best-fit values. Dashed vertical lines further denote the marginalized $1\sigma$ intervals for each parameter, with blue for FIM and red for MCMC. In the bottom-right panel showing the $(\log_{10}\Omega_0, \log_{10} (f_p / \rm Hz))$ plane, dot-dashed and dashed curves indicate the 1$\sigma$ and 2$\sigma$ confidence intervals, respectively, with black for the FIM and green for MCMC. This provides a more detailed comparison between the two approaches and lays the groundwork for subsequent studies on parameter constraints. The inset displays the signal spectra and detector sensitivities. The latter characterizes the strength of the instrumental noise and thus indicates the detectability of the experiment, including contributions from both the astrophysical and phase-transition SGWB components.
    }
    \label{yff}  
\end{figure}

In contrast, the MCMC method offers a more robust and flexible framework that explores the full posterior landscape without assuming any specific distributional form. It can accurately characterize non-Gaussian features, parameter degeneracies, and multimodal distributions, albeit at the cost of higher computational demands.

Figure~\ref{yfd} presents a concrete example of such a comparison. We focus on four representative parameters in the SGWB model: two related to instrumental noises, $N_{\rm acc}$ and $\delta x$, and two associated with the astrophysical background, $\Omega_{\rm ast}$ and $\varepsilon$. The injected values are $N_{\rm acc} = 3 \times 10^{-15}$, $\delta x = 8 \times 10^{-12}$, $\log_{10} \Omega_{\rm ast} = -11.5$, and $\varepsilon = 2/3$, indicated by the blue lines, while the red lines indicate the parameter values recovered from the MCMC analysis, corresponding to a SNR of 739. The 68\% and 95\% confidence regions predicted by the FIM are shown as dark and light blue ellipses, respectively, and the corresponding credible regions from MCMC sampling are displayed in red. In addition, the $1\sigma$ ranges from the marginalized one-dimensional posteriors are indicated by dashed lines, with blue for FIM and red for MCMC. From these plots, we find that the relative uncertainties estimated via the FIM, using the bound in Eq.~\eqref{STE}, are approximately 0.088\% for $N_{\rm acc}$, 0.006\% for $\delta x$, 0.033\% for $\log_{10}\Omega_{\rm ast}$, and 0.738\% for $\varepsilon$. The corresponding uncertainties from the MCMC posteriors are nearly identical; 0.088\%, 0.006\%, 0.033\%, and 0.728\%. These results demonstrate high precision in parameter recovery across both approaches.

% \FloatBarrier

\subsection{Measurements of the thermodynamics parameters}

Here, in addition to the contributions from instrumental noises and the astrophysical SGWB, we incorporate the SGWB generated by the EWPT, specifically the dominant contribution from SWs discussed earlier. The likelihood function remains the same as in previous analyzes, with a modified signal model to include this additional component,

\begin{equation}\label{sml}
    S_{A} = S_{E} =
    \frac{3 H_{0}^{2}}{4 \pi^{2}} 
    \frac{\Omega_\text{ast}\left(\tfrac{f}{f_{\text{ref}}}\right)^{\varepsilon} + \Omega_{\text{sw}}(f)}{f^{3}} \, \mathcal{R}_A \;,
\end{equation}
where $\Omega_{\text{sw}}(f)$ is given by Eq.~\eqref{pgv} and is characterized by two parameters; the peak amplitude $\Omega_0$ and the peak frequency $f_{\mathrm{p}}$.

Using this extended signal model, we estimate the joint posterior distribution for six parameters; two instrumental noise parameters, $N_{\text{acc}}$ and $\delta x$, two astrophysical background parameters, $\Omega_{\text{ast}}$ and $\varepsilon$, and two parameters characterizing the phase transition of the SGWB, $\Omega_0$ and $f_{\mathrm{p}}$. The resulting confidence ellipses are shown in Fig.~\ref{yff}, providing a comprehensive view of the parameter correlations and uncertainties as derived from both the FIM and full posterior exploration via MCMC sampling within the Bayesian framework. This comparison allows us to assess the validity of the Gaussian approximation in a higher-dimensional, physically motivated parameter space. The injected values, used consistently in both simulation and inference, are $N_{\text{acc}} = 3 \times 10^{-15}$, $\delta x = 8 \times 10^{-12}$, $\log_{10} \Omega_{\rm ast} = -11.5$, $\varepsilon = 2/3$, $\log_{10} \Omega_0 = -12.0$, and $\log_{10} (f_p / \rm Hz) = -2.13$. With this set of parameters, the calculated SNRs for the astrophysical background and the SW contribution are 739 and 52, respectively. The relative uncertainties predicted by the FIM for these parameters are approximately 0.096\%, 0.007\%, 0.057\%, 1.333\%, 0.491\%, and 1.29\%, respectively. In contrast, the corresponding uncertainties estimated from MCMC sampling are very similar; 0.097\%, 0.007\%, 0.057\%, 1.329\%, 0.552\%, and 1.406\%.

\FloatBarrier

At first glance, the MCMC-derived relative uncertainty for $\varepsilon$ appear slightly smaller than those predicted by the FIM, which may seem counterintuitive since the FIM is generally expected to provide the best forecasts under the assumption of a locally Gaussian posterior. This apparent tension can be understood by recalling that Fisher forecasts probe only the local curvature of the likelihood around the maximum-likelihood point. By construction, they neglect non-Gaussian features that are naturally captured in MCMC analyzes. In practice, the confidence ellipses from the FIM primarily illustrate the local shape of the posterior, whereas the standard deviations derived from MCMC chains account for the full posterior volume, including possible asymmetries, long tails, and mild nonlinearities. As a result, the projected ellipses from MCMC can appear comparable or even slightly tighter, while the corresponding marginalized uncertainties remain broader. The inclusion of $\Omega_0$ and $f_{\mathrm{p}}$ adds further structure to the likelihood surface since these parameters directly determine the amplitude and peak frequency of the phase-transition signal. Their coupling with $\Omega_{\text{ast}}$ and $\varepsilon$ introduces moderate degeneracies, which broaden the confidence regions along certain directions in parameter space. Overall, while the FIM remains a valuable first-order approximation, its local Gaussian nature becomes less reliable as dimensionality and model complexity increase, underscoring the need for full Bayesian inference to obtain robust uncertainty quantification.

\begin{figure}[t]
    \centering
    \includegraphics[width=0.7\textwidth]{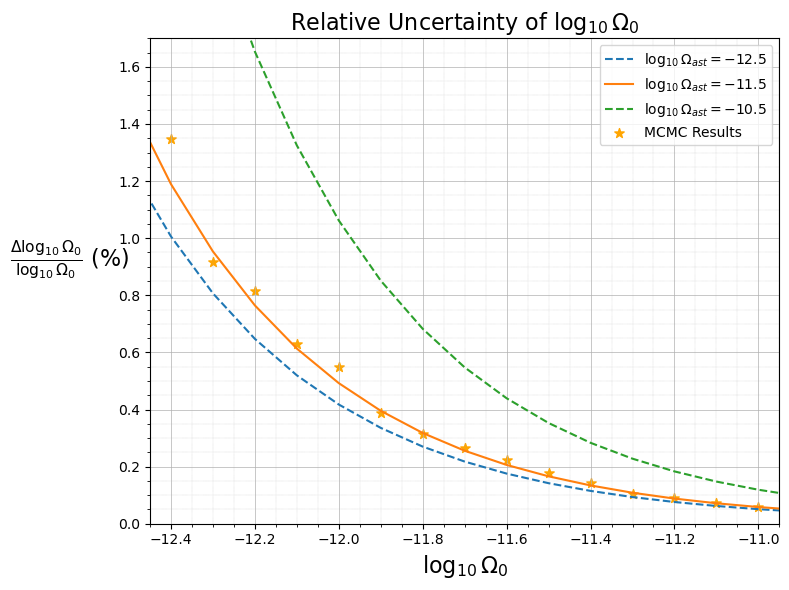}
    \caption{
    Relative uncertainty of $\log_{10} \Omega_0$, defined as $\Delta \log_{10} \Omega_0 / \log_{10} \Omega_0$, as a function of $\log_{10} \Omega_0$ for different astrophysical background amplitudes $\log_{10} \Omega_{\text{ast}}$, derived using the FIM (lines) and MCMC for one scenario (points). The MCMC results (orange points) are consistent with the Fisher forecasts (orange line), both obtained for $\log_{10} \Omega_{\text{ast}}=-11.5$. 
    }
    \label{yef}  
\end{figure}

To further investigate how the detectability of the phase transition signal depends on the fiducial model parameters, we analyze the relative uncertainty of $\log_{10} \Omega_0$ as a function of its injected value. In particular, we also explore how this uncertainty is influenced by the presence of an overlapping astrophysical SGWB component, characterized by different values of $\log_{10} \Omega_{\text{ast}}$. Using the FIM formalism, we compute the relative error $\frac{\Delta \log_{10} \Omega_0}{\log_{10} \Omega_0}$ under three representative scenarios: $\log_{10} \Omega_{\text{ast}} = -10.5$, $-11.5$, and $-12.5$. The results, illustrated in Fig.~\ref{yef}, demonstrate that a stronger astrophysical background increases the uncertainty in $\Omega_0$, especially when $\Omega_0$ is small, as expected. This highlights the importance of disentangling different SGWB components when interpreting observational data. We have also shown in this figure the comparison between MCMC-derived uncertainties (with stars) and those from the FIM for one case, which demonstrates the overall agreement between the two approaches.

\begin{figure}[t]
    \centering
    \includegraphics[width=0.9\textwidth]{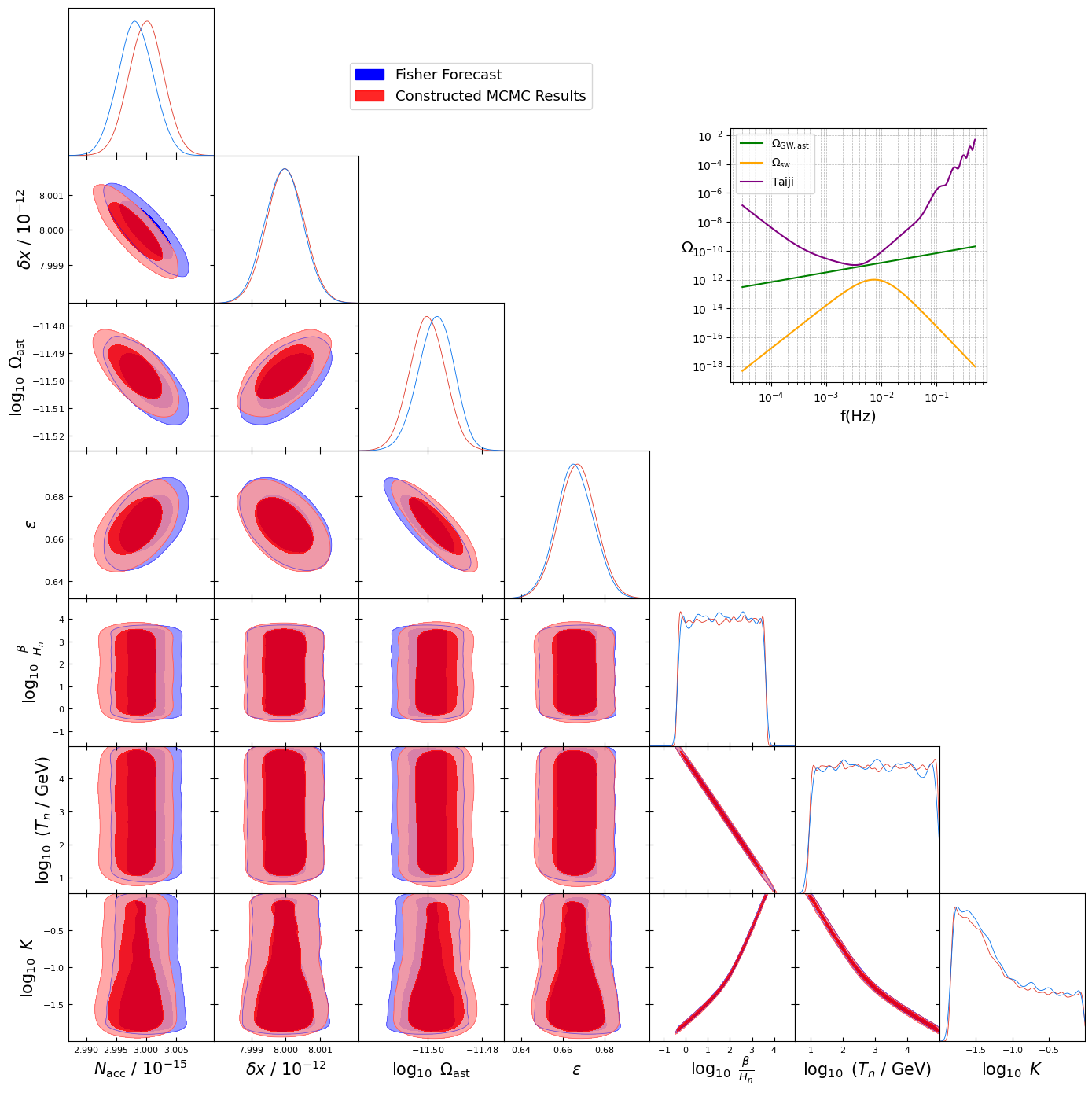}
    \caption{
    Confidence ellipses for seven model parameters; two instrumental noise parameters, $N_{\rm acc}$ and $\delta x$, two astrophysical parameters, $\Omega_{\rm ast}$ and $\varepsilon$, and three thermodynamic parameters of the EWPT, $\beta/H_{\rm n}$, $T_\mathrm{n}$, and $K = \alpha/(1+\alpha)$. The dark and light blue areas denote the 68\% and 95\% confidence regions predicted by the FIM, respectively, while the red contours show the corresponding credible regions from MCMC sampling. To reduce degeneracies in the thermodynamic description, $T_\mathrm{n}$ is fixed at random values drawn from a uniform distribution, allowing for clearer resolution of the remaining parameter correlations. The comparison reveals both the degeneracies among thermodynamic parameters and the extent to which the Fisher approximation captures the posterior structure. 
    }
    \label{fff}  
\end{figure}

In the case of the full SWs model described by Eq.~\eqref{tdv}, parameter degeneracies naturally arise from the presence of three free thermodynamic parameters: $\beta / H_\mathrm{n}$, $T_\mathrm{n}$, and $K = \alpha / (1 + \alpha)$. These degeneracies pose challenges for accurate parameter inference and interpretation. To mitigate this issue, we follow the strategy of \cite{Caprini:2024hue}, fixing $T_\mathrm{n}$ at random values drawn from a uniform distribution. The resulting confidence ellipses for all parameters are shown in Fig.~\ref{fff}, enabling a direct comparison between Fisher forecasts and posterior distributions. This approach provides insight into the attainable parameter constraints and shows the impact of degeneracies under the thermodynamic parametrization. Our results can be compared with the analysis of \cite{Caprini:2024hue}. 

To summarize, the FIM provides a fast and computationally efficient way to forecast uncertainties under the assumption of a local Gaussian posterior. In contrast, MCMC sampling—though computationally more demanding—yields the full posterior distribution and faithfully captures non-Gaussian features and parameter degeneracies. In both cases, we observe overall agreement between the two approaches, and have thus obtained the measurement uncertainties for the parameters governing the GW spectrum and that of the phase transitions.

\subsection{Measurements of xSM model parameters} 

\begin{figure}[t]
    \centering
    \includegraphics[width=0.418\linewidth]{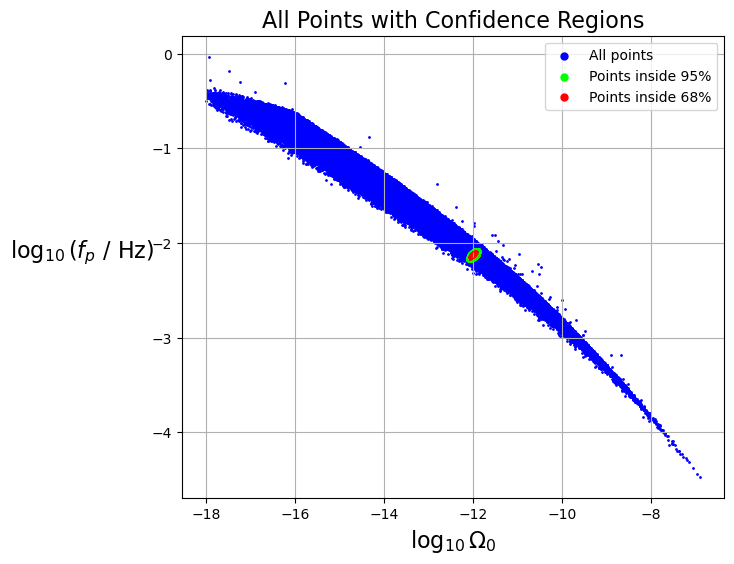}
    \quad
    \includegraphics[width=0.45\linewidth]{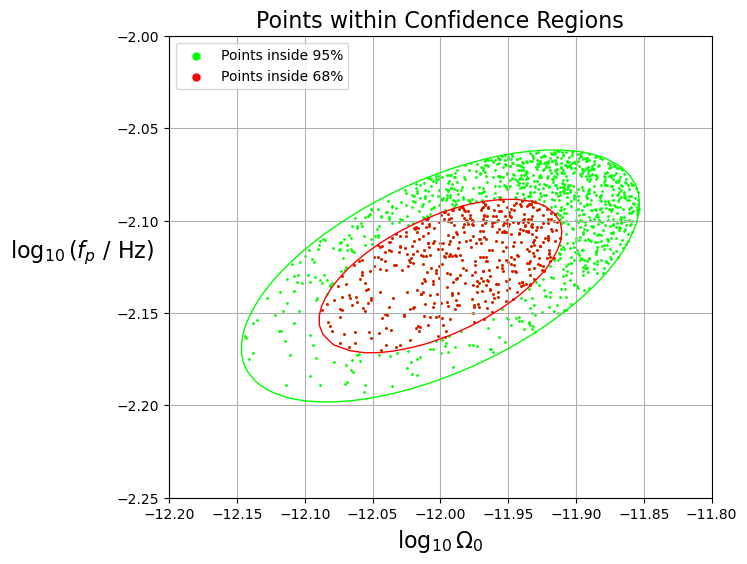}
    \\
    \includegraphics[width=0.422\linewidth]{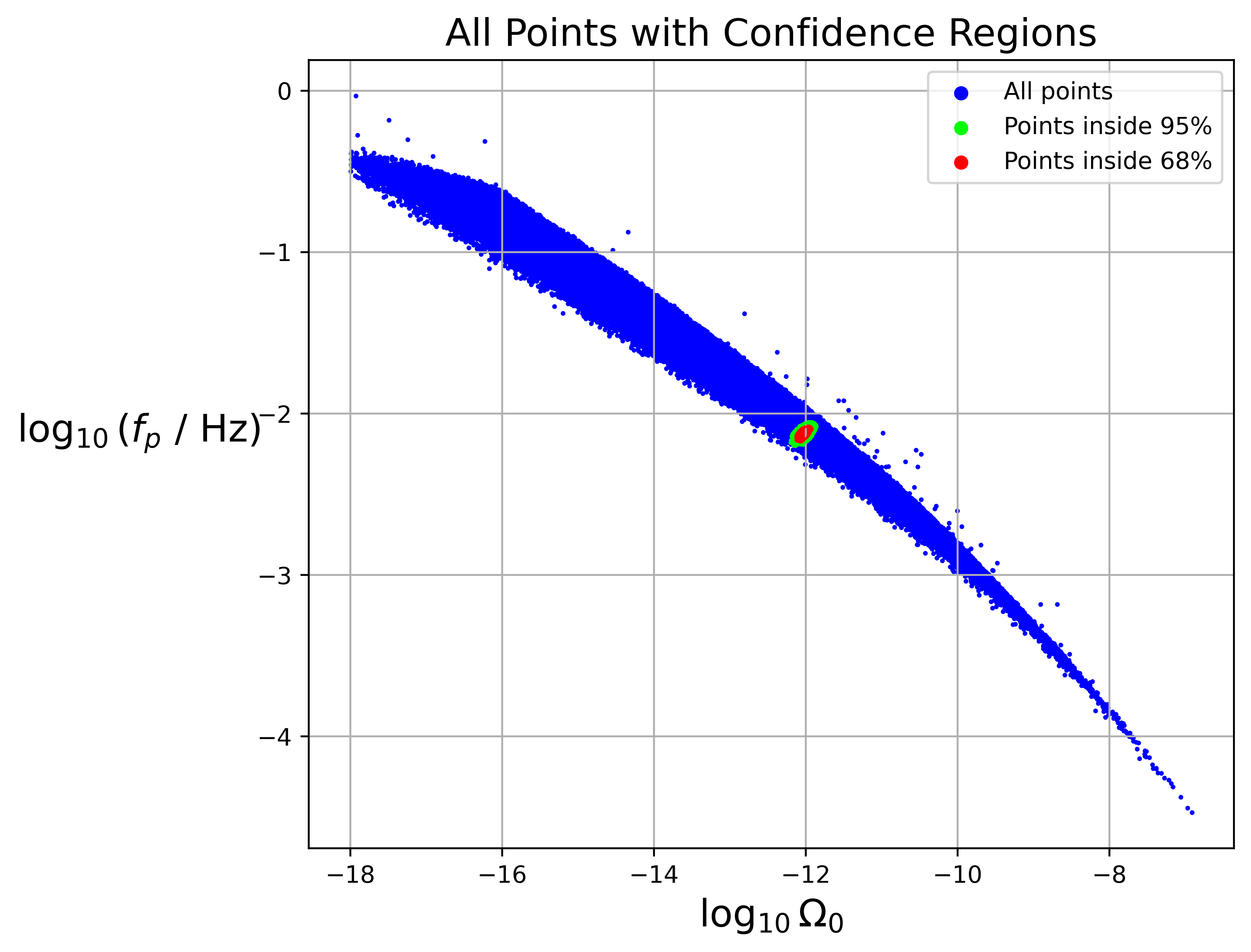}
    \quad
    \includegraphics[width=0.45\linewidth]{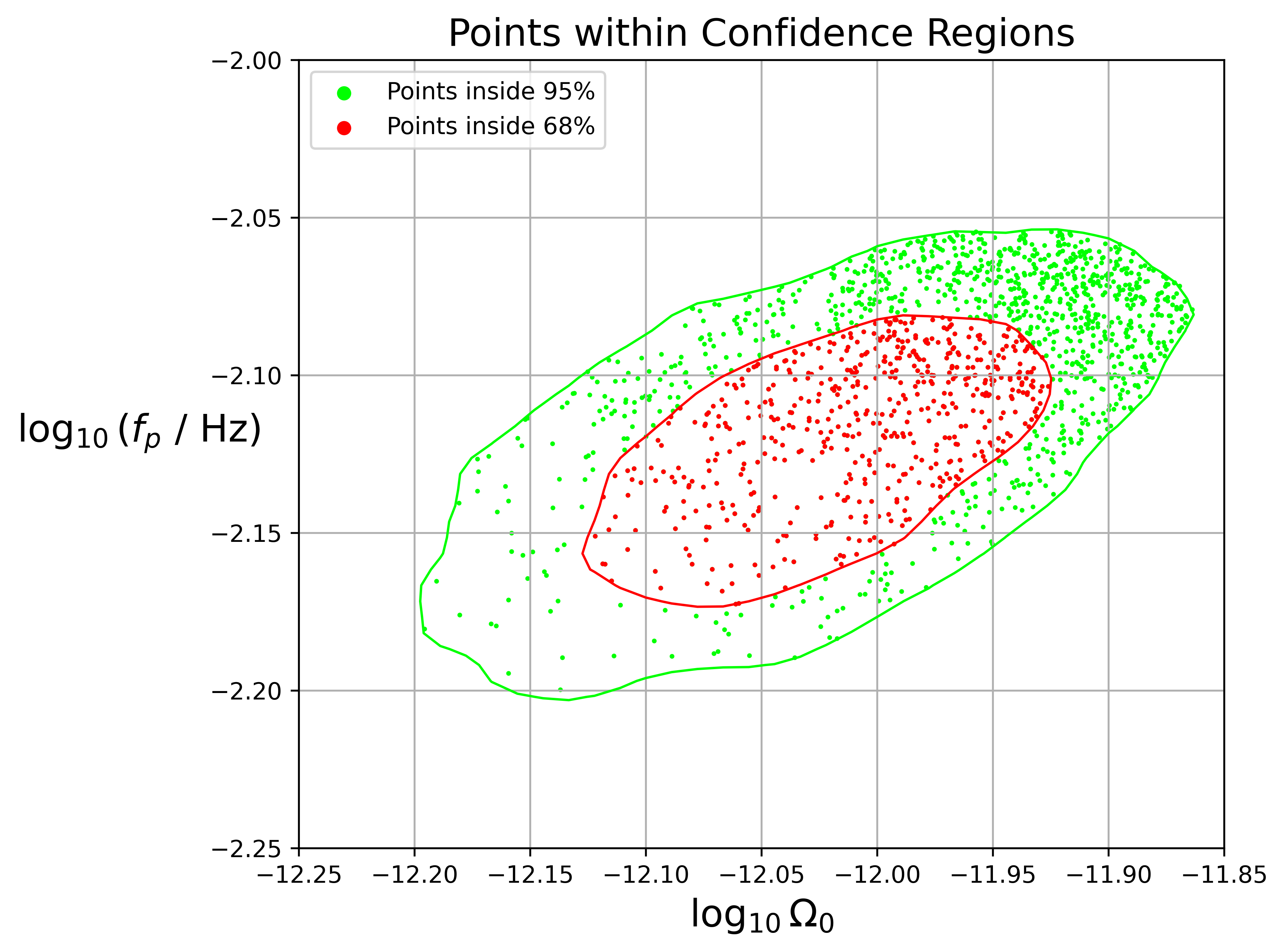}
    \caption{
    Projection of the scanned xSM model points onto the spectral parameter plane \(\log_{10} \Omega_0\)–\(\log_{10} (f_p / \rm Hz)\). The top row corresponds to FIM forecasts, while the bottom row shows results from MCMC sampling. The left panels display scanned parameter points in the geometric parameter space, with blue dots representing the full xSM dataset that survives the phenomenological and
    theoretical constraints and is capable of generating a first order EWPT. The right panels provide a zoomed-in view of the region surrounding the benchmark prediction. Points falling within the 68\% and 95\% confidence regions are highlighted in red and green, respectively. This visualization enables a direct comparison between Fisher-based and MCMC-based constraints on GW spectral parameters illustrating how viable model points are distributed with respect to the forecasted confidence contours.
    }
    \label{uem}
\end{figure}

Due to the aforementioned degeneracy, we adopt a strategy based on the directly inferred spectral parameters $(\log_{10}\Omega_0,\log_{10} (f_p / \rm Hz))$ and identify the regions in the xSM parameter space that are consistent with these measurements. To illustrate how the resulting constraints propagate into the underlying particle-physics model, we project the scanned xSM parameter points onto the spectral parameter plane $(\log_{10}\Omega_0,\log_{10} (f_p / \rm Hz))$. Figure~\ref{uem} presents this projection for both FIM forecasts and MCMC sampling, showing the full set of viable xSM points as well as those lying within the 68\% and 95\% confidence regions around the benchmark prediction, thereby enabling a direct comparison between FIM-based and MCMC-based constraints.

% \FloatBarrier

We then map the constrained regions of the spectral parameters to the parameter space of the xSM, characterized by the five parameters \((v_s, m_{h_2}, \theta, b_3, b_4)\). These regions are further restricted through an extensive scan that ensures compliance with all phenomenological and theoretical constraints. For each point in this viable parameter space, we calculate the phase–transition parameters \((\alpha, \beta/H_{n}, T_{n})\) and correspondingly \((\Omega_0, f_{\mathrm{p}})\). If the prediction lies within the FIM/MCMC acceptance bands, we will keep that point. Projecting the accepted points back onto \((v_s, m_{h_2}, \theta, b_3, b_4)\) shows which parameter values are allowed and how they are related, as shown in Fig.~\ref{qmd}. As the constraints from the FIM and the MCMC method are similar, we will now, and in the following discussion, show only the more precise results obtained from the MCMC. 

From these plots, we can see that the constrained region on the spectral parameters now leads to a similarly constrained parameter space for the particle physics model. We also note the apparent impact of the problem of parameter degeneracy, which results in a more spread-out feature of the $68\%$ and $95\%$ regions. This is obvious, as the mapping
\[  (v_s, m_{h_2}, \theta, b_3, b_4) \mapsto (\Omega_0, f_{\mathrm{p}})  \]
is not unique. Reducing from five parameters to two always creates degeneracies. This means that the FIM and MCMC results do not select a single point; instead, they provide bands or islands of possible parameter values that can produce the same GW spectrum. The projections in Fig.~\ref{qmd} show these degeneracies clearly. These GW-selected regions provide a compact, data-driven summary of the xSM parameter space that is consistent with a potential detection. 

\begin{figure}[t]
    \centering
    \includegraphics[width=0.8\linewidth]{figure/xsm_mcmc_filter.png}
    \caption{
    Corner plot showing the distribution of five xSM model parameters—\(v_s\), \(m_{h_2}\), \(\theta\), \(b_3\), and \(b_4\)—corresponding to the MCMC-measured confidence regions in the spectral parameter plane \((\log_{10} \Omega_0,\, \log_{10} (f_p / \rm Hz))\), as previously displayed in Fig.~\ref{uem}. Each panel presents the pairwise correlations between two model parameters (off-diagonal) or the marginalized one-dimensional distribution of a single parameter (diagonal). Blue points represent the entire ensemble of scanned xSM parameter sets that survive phenomenological and theoretical constraints and can produce GW signals. Red and green points indicate the subsets that lie within the MCMC-derived 68\% and 95\% confidence regions, respectively, in the spectral space. These subsets are here projected back into the model parameter space to examine how observational constraints influence the viable ranges of theoretical inputs. 
    }
    \label{qmd}
\end{figure}

\begin{figure}[t]
    \centering
    \includegraphics[width=0.9\linewidth]{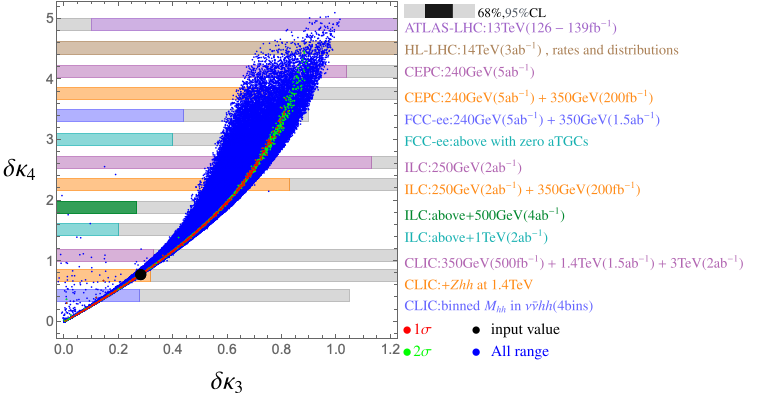}
    \caption{
     Inferred Higgs cubic and quartic couplings \(\delta\kappa_3\) and \(\delta\kappa_4\) based on GW measurement of the simulated benchmark signal and MCMC sampling. Blue points represent the full parameter space that survive phenomenological and theoretical constraints and which also can lead to first order EWPT; red and green points highlight subsets that fall within the \(1\sigma\) and \(2\sigma\) confidence regions, respectively, in the \((\log_{10} \Omega_0,\, \log_{10} (f_p / \rm Hz))\) plane. The black dot denotes the prediction from the fiducial benchmark point. Overlaid horizontal bars indicate the projected \(68\%\) and \(95\%\) confidence level sensitivities to \(\delta\kappa_3\) at various future collider experiments, including HL-LHC, CEPC, ILC, FCC, and CLIC taken from~\cite{DiVita:2017vrr}. 
     }
    \label{psm}
\end{figure}

\subsection{Measurements of Higgs self-couplings}

With the xSM parameter range obtained from the GW measurement of the simulated signal, we can take one step further and make predictions for various observables that may or may not be measured at colliders and other experiments. One particularly important set of observables of this kind is the Higgs self-couplings, or equivalently, deviations of the Higgs self-couplings, \(\delta\kappa_3\) and \(\delta\kappa_4\), from their SM values, as defined in Eq.~\eqref{hdb}. These two couplings describe the shape of the scalar potential and are therefore important for understanding electroweak symmetry breaking and for exploring possible extensions of the Higgs sector. 

%\begin{figure}[H]
%    \centering
%    \includegraphics[width=0.9\linewidth]{figure/xsm_fisher_filter.png}
%    \caption{
%    Corner plot showing the distribution of five xSM model parameters—\(v_s\), \(m_{h_2}\), \(\theta\), \(b_3\), and \(b_4\)—corresponding to the Fisher-forecasted confidence regions in the spectral parameter plane \((\log_{10} \Omega_0,\, \log_{10} (f_p / \rm Hz))\), as previously displayed in Fig.~\ref{uem}. Each panel presents the pairwise correlations between two model parameters (off-diagonal) or the marginalized one-dimensional distribution of a single parameter (diagonal). Blue points represent the full set of scanned xSM parameter combinations that yield viable GW signals. Among them, red and green points denote those that lie within the Fisher-predicted 68\% and 95\% confidence regions, respectively, in the spectral parameter space. These colored subsets have been projected back into the original model parameter space to assess the impact of GW spectral constraints on theoretical model parameters. This figure provides a detailed visualization of how precision measurements of GW spectral features can constrain the fundamental parameters governing electroweak symmetry breaking in the extended scalar sector.
%    }
%    \label{udc}
%\end{figure}

Measuring them directly at colliders is, however, very challenging. The quartic coupling \(\delta\kappa_4\) is beyond the reach of planned detectors~\cite{ATLAS:2024xcs}, while the trilinear coupling \(\delta\kappa_3\) is extracted from the measurement of double Higgs production in various channels. The latest measurement was recently performed by ATLAS in the channel $b \bar{b} \gamma\gamma$, using the full data from Run 2 and a portion from Run 3, with a total integrated luminosity of $308\ \text{fb}^{-1}$ at $\sqrt{s} = (13, 13.6) \text{TeV}$. The constraint thus obtained is $-2.7 < \delta \kappa_{\lambda} < 5.6$ at $95\%\ \text{CL}$~\cite{ATLAS:2025nda}. For future colliders, the expectation for the measured trilinear coupling is shown by the color bands in Fig.~\ref{psm}, taken from~\cite{DiVita:2017vrr}.\footnote{Since there is no sensitivity here to the quartic coupling $\delta \kappa_4$, the vertical positioning of these bands is solely for illustrative purposes, with no connection to the values of $\delta \kappa_4$.} The experimental determination of the cubic coupling is quite uncertain at this moment; however, while it can be improved in the future, a precise determination of the quartic coupling $\delta \kappa_4$ would still be very difficult in the long term.

% \FloatBarrier

The question naturally arises as to whether the GW measurement from EWPT can be used to provide a better determination of these two couplings. To do this, from the parameter space of the xSM identified from the GW measurement, we infer the corresponding $\delta \kappa_3$ and $\delta \kappa_4$. These values are added onto Fig.~\ref{psm}, where the blue points denote all the points that can give a first-order EWPT and survive phenomenological constraints. The green and red points denote those that provide the GW measurement within one and two standard deviations, respectively, as in previous sections. In this way, we are actually combining future GW measurements with current collider and other phenomenological and theoretical constraints.

For all points, due to the features of this model, as explained in more detail in~\cite{Alves:2018oct}, there appears to be a linear correlation between these two couplings, which can be more easily understood from a Taylor expansion in powers of the mixing angle $\theta$, taking into account the concentrated regions of the parameter space resulting from all phenomenological and theoretical constraints. Comparing the blue points with the green and red points, we can see that the region on the plane $(\delta \kappa_3, \delta \kappa_4)$ has narrowed down to a much smaller area due to the GW measurement. This highlights the importance of GW measurement in probing the Higgs couplings, especially regarding the quartic ones. We note, however, that due to the same parameter degeneracy problem, the precision on $\delta \kappa_3$ and $\delta \kappa_4$ is limited. This limitation, at least for this model, could potentially affect a much broader class of models. A potential solution to this conundrum is to make use of finer structures on the spectrum, such as the damping feature at higher frequencies caused by dissipative effects in the fluid~\cite{Guo:2023koq}. We further note that in deriving the results here, we have neglected theoretical uncertainties arising from spectrum and phase transition parameter calculations, as explored recently~\cite{Lewicki:2024xan}. The inclusion of these would lead to a less constraining result. Systematically quantifying and including these uncertainties in the Bayesian framework, along with a more faithful detector simulation, would be a direction for future studies.

\section{Conclusion}\label{kjh}

In this work, we have developed a comprehensive framework that connects the theoretical modeling of FOPTs in the xSM with realistic data analysis strategies for space-based GW detectors, such as Taiji and LISA. By combining frequency-domain detector response modeling, astrophysical foregrounds, and instrumental noise, we constructed simulated datasets for the SGWB and carried out both FIM and Bayesian MCMC analyzes. Using simulated data, we first estimated instrumental and astrophysical parameters, and then extended the analysis to include the stochastic signal from SWs generated during a first-order EWPT. The inferred spectral parameters $(\Omega_0, f_{\mathrm{p}})$ were subsequently mapped onto the xSM parameter space $(v_s, m_{h_2}, \theta, b_3, b_4)$, enabling us to identify viable regions consistent with a potential SGWB detection. This mapping further allowed us to translate GW constraints into predictions for Higgs self-coupling deviations $(\delta\kappa_3, \delta\kappa_4)$, which encode the shape of the electroweak scalar potential. Our analysis shows that space-based GW observations can help determine these couplings. While collider measurements of $\delta\kappa_3$ and especially $\delta\kappa_4$ remain extremely challenging, GW-based inference from FOPTs offers an indirect but powerful probe of scalar self-interactions. The synergy between collider physics and GW astronomy thus opens a new pathway toward a more complete understanding of electroweak symmetry breaking and the origin of the Higgs potential. Future work will extend this framework to include detector networks, realistic data gaps, and a joint analysis within the global-fit approach that simultaneously accounts for deterministic sources such as massive black-hole binaries. The inclusion of theoretical uncertainties in the phase-transition modeling—such as gauge dependence and finite-temperature corrections—will also be an important next step toward robust, data-driven constraints on beyond-Standard-Model physics from the upcoming era of space-based GW observations.

\section*{Acknowledgments}
We would like to thank Ligong Bian, Ju Chen, Ming-Hui Du, Chang Liu, and Michael J. Ramsey-Musolf for helpful discussions. This work is supported by the startup fund provided by the University of Chinese Academy of Sciences and by the National Natural Science Foundation of China (NSFC) under Grants No. 12335005, No. 12475109, and No. 12547104.

\section*{Data availability}
The data that support the findings of this article are not publicly available upon publication because it is not technically feasible and/or the cost of preparing, depositing, and hosting the data would be prohibitive within the terms of this research project. The data are available from the authors upon reasonable request.

\FloatBarrier
\appendix
\section{Fisher Iinformation matrix and Bayesian inference}\label{app:fim}

In this appendix, we provide a detailed introduction to the FIM and its formal derivation, followed by a description of the Bayesian inference framework adopted in this work. To illustrate and compare these statistical approaches in a transparent setting, we further present a simple linear toy model as an explicit example of the parameter-inference procedure.

%%%%%%%%%%%%%%%%%%%%%%%%%%%%%%%%%%%%%%%%%%%%%%%%%%%%%
\begin{figure}[t]
    \centering
    \includegraphics[width=1\textwidth]{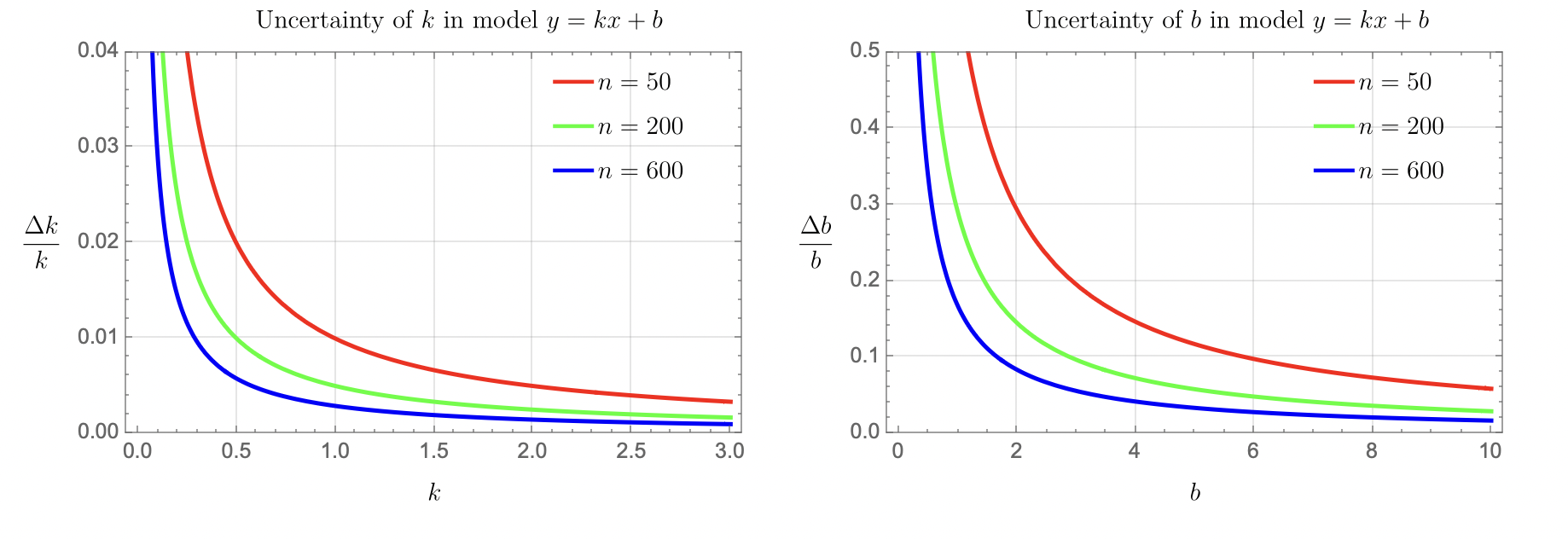}
    \caption{
    Uncertainty analysis for the parameters $k$ (left) and $b$ (right) in the linear model $y = kx + b$ based on the FIM. The figure compares the relative uncertainties for three different values of the data size $n$. As expected, a larger number of data points leads to smaller parameter uncertainties.
    }
    \label{fig:1}  
\end{figure}
%%%%%%%%%%%%%%%%%%%%%%%%%%%%%%%%%%%%%%%%%%%%%%%%%%%%%

\subsection{Fisher information matrix and the Cramér-Rao lower bound} 

In this section, we derive the general expression for the FIM. Before doing so, it is necessary to examine the analytical form of the log-posterior function~\cite{Gregory_2005}, which forms the basis for the subsequent derivation,

\begin{equation}\label{skb}
    \ln \left[p(\boldsymbol{\theta})\mathcal{L}(\boldsymbol{\theta})\right]
    =
    \ln \left[p(\hat{\boldsymbol{\theta}})\mathcal{L}(\hat{\boldsymbol{\theta}})\right]
    -\frac{1}{2}\sum_{ij} 
    \left(\theta_i-\hat\theta_i\right) (C^{-1})_{ij} \left(\theta_j-\hat\theta_j\right) ,
\end{equation}
where $\boldsymbol{\hat{\theta}}$ denotes the maximum likelihood estimate of the parameters, $\boldsymbol{C}$ represents the covariance matrix, $\theta_i$ refers to the $i$ th parameter, and $p(\boldsymbol{\theta})$ is the prior distribution (as mentioned above, uniform priors are adopted for all model parameters). The inverse of the covariance matrix, $\boldsymbol{C}^{-1}$, is defined as the FIM, which is a matrix of second-order derivatives of $\ln \!\left[p(\boldsymbol{\theta}) \mathcal{L}(\boldsymbol{\theta})\right]$ evaluated at $\boldsymbol{\theta} = \hat{\boldsymbol{\theta}}$,
\begin{equation}\label{urc}
    F_{ij} \equiv (C^{-1})_{ij} =
    -E 
    \left(
        \frac{\partial^2 \ln p(\boldsymbol{\theta})\mathcal{L}(\boldsymbol{\theta})}{\partial\theta_i\partial\theta_j} 
    \right) ,
\end{equation}
where $E$ denotes the expectation value. It is useful to divide it into two parts,
\begin{equation}
    F_{ij}=F_{ij}^\text{prior}+F_{ij}^\text{likelihood} ,
\end{equation}
where
\begin{equation} \label{wmd}
\begin{split}
    F_{ij}^\text{prior} &=
    -E 
    \left(
        \frac{\partial^2 \ln p(\boldsymbol{\theta})}{\partial\theta_i\partial\theta_j} 
    \right)
    ,\qquad
    \\
    F_{ij}^\text{likelihood} &=
    -E 
    \left(
        \frac{\partial^2 \ln \mathcal{L}(\boldsymbol{\theta})}{\partial\theta_i\partial\theta_j} 
    \right) .
\end{split}
\end{equation}
For an unbiased estimator $\hat{\theta}_i$ of an unknown parameter, the CRLB imposes the following lower bound on its standard deviation \cite{doi:https://doi.org/10.1002/9780470173862.ch4}:
\begin{equation}\label{STE}
    \Delta \hat\theta_i\geq \sqrt{(F^{-1})_{ii} } .
\end{equation}  

Thus, the relative uncertainty associated with an estimator $\hat{\theta}_i$ is defined as $\tfrac{\Delta \hat{\theta}_i}{\hat{\theta}_i}$. According to the CRLB, the inverse of the FIM provides critical insights into the precision of parameter estimation. Specifically, the diagonal elements of the inverse FIM, $(F^{-1})_{ii}$, correspond to the minimum achievable variances of unbiased estimators for the parameters $\theta_i$. These values define the theoretical lower bounds on the uncertainties associated with each parameter. In contrast, the off-diagonal elements quantify the covariances between parameter pairs, capturing the extent of their statistical correlation or dependency.

To visualize these correlations, one may construct confidence ellipses, which provide a geometric interpretation of the joint uncertainty between two parameters. These ellipses are derived from the second term in Eq.~(\ref{skb}). The construction begins by analyzing the so-called Mahalanobis distance \cite{Mahalanobis:1936gd},

\begin{equation}
    d_M^2 
    = \sum\limits_{ij}
    \left(\theta_i-\hat\theta_i\right) (C^{-1})_{ij} \left(\theta_j-\hat\theta_j\right) ,
\end{equation}
which is defined in a multidimensional space.

In the specific case of constructing confidence ellipses, we focus on a two-parameter subspace with parameters $\theta_\alpha$ and $\theta_\beta$. The associated covariance matrix, denoted by $\Sigma$, is constructed from the components of $C$,
\begin{equation}
    \Sigma = 
    \begin{pmatrix}
        C_{\alpha\alpha} & C_{\alpha\beta} \\
        C_{\alpha\beta} & C_{\beta\beta} 
    \end{pmatrix} ,
\end{equation}
and can be diagonalized through the eigendecomposition,
\[
\Sigma = U \Lambda U^T,
\]
where $\Lambda$ is a diagonal matrix whose entries are the eigenvalues of $\Sigma$, and $U$ is the orthogonal matrix of the corresponding eigenvectors.  

We define the matrix square root of the inverse covariance matrix as
\[
\Sigma^{-\frac{1}{2}} = U \Lambda^{-\frac{1}{2}} U^T,
\]
and construct a pair of normalized parameters,
\begin{equation}
    \begin{pmatrix}
        z_\alpha \\
        z_\beta
        \end{pmatrix}
        =
        \Sigma^{-\frac{1}{2}}
        \begin{pmatrix}
        \theta_\alpha - \hat{\theta}_\alpha \\
        \theta_\beta - \hat{\theta}_\beta
    \end{pmatrix},
\end{equation}
which follows a standard normal distribution. Under this transformation, the Mahalanobis distance in the two-parameter subspace simplifies to the following form:

\begin{equation}
d_M^2 = z_\alpha^2 + z_\beta^2.
\end{equation}
The Mahalanobis distance, $d_M^2$, follows a chi-squared distribution with degrees of freedom equal to the number of parameters under consideration. In the case of constructing confidence ellipses for two parameters, the distribution is $\chi^2(n=2)$. Accordingly, the confidence ellipse represents the region in which the joint probability density satisfies
\[
d_M^2 = c,
\]
where $c$ is a constant determined by the desired confidence level. For instance, $c = 2.30$ corresponds to a 68\% confidence region, and $c = 6.18$ corresponds to a 95\% confidence region.

The resulting confidence ellipse visually illustrates the uncertainties in the two parameters and their mutual correlation. A highly elongated ellipse indicates a strong correlation between the parameters, whereas a circular shape implies a weak or no correlation. The shape and orientation of the ellipse thus offer valuable insights into the coupling and degeneracy structure of the parameter space, as well as the overall precision of the parameter estimation.

In summary, the inverse FIM encapsulates both the variances and covariances of the model parameters, forming the foundation for constructing confidence regions. The confidence ellipse, grounded in the Gaussian approximation of the likelihood and chi-squared statistics, serves as an effective and interpretable tool for visualizing joint parameter uncertainties and correlations.

\subsection{Bayesian inference and MCMC sampling}

Bayesian inference provides a coherent framework for parameter estimation and hypothesis testing, grounded in the principles of probability theory. At its core lies Bayes' theorem, which relates the posterior probability distribution of model parameters to the prior knowledge and the likelihood of the observed data,
\begin{equation}
    p(\boldsymbol{\theta} \mid \boldsymbol{d}, \mathcal{M}) = \frac{p(\boldsymbol{d} \mid \boldsymbol{\theta}, \mathcal{M}) \, p(\boldsymbol{\theta} \mid \mathcal{M})}{p(\boldsymbol{d} \mid \mathcal{M})},
\end{equation}
where $\boldsymbol{\theta}$ denotes the set of model parameters, $\boldsymbol{d}$ represents the observed data, and $\mathcal{M}$ denotes the underlying model. The term $p(\boldsymbol{d} \mid \boldsymbol{\theta}, \mathcal{M})$ is the likelihood function, which quantifies the probability of observing the data given a specific choice of parameters. The prior distribution $p(\boldsymbol{\theta} \mid \mathcal{M})$ encodes preexisting knowledge or assumptions about the parameters before any data are considered. The denominator, $p(\boldsymbol{d} \mid \mathcal{M})$, is the evidence or marginal likelihood, which acts as a normalization constant and plays a central role in model comparison.

In most practical applications, particularly when dealing with high-dimensional or nonlinear models, the posterior distribution cannot be easily determined. To address this, numerical techniques are employed, among which MCMC methods are the most widely used. In this work, we perform MCMC sampling using the \texttt{PyMC} probabilistic programming framework~\cite{AbrilPla:2023pymc}, which provides efficient implementations of state-of-the-art samplers such as Metropolis and NUTS. After an initial burn-in phase, the chains are expected to explore the parameter space adequately in a manner proportional to the posterior probability density.

These methods enable the estimation of summary statistics of the posterior, such as the mean, median, credible intervals, and correlation structures between parameters. Despite its robustness, MCMC sampling can be computationally intensive, especially for complex likelihood functions or large datasets. Convergence diagnostics and autocorrelation analyzes are crucial for ensuring the validity of the results. Nevertheless, when properly applied, MCMC provides an indispensable tool for Bayesian inference in both theoretical modeling and data-driven investigations.

\subsection{A toy model: \texorpdfstring{$y=kx+b$}{y=kx+b}}

In this subsection, we compare the parameter uncertainties derived from the two approaches mentioned above. The first is based on the CRLB, which provides a theoretical minimum variance for any unbiased estimator and is computed using the FIM. The second approach employs MCMC sampling to characterize the full posterior distribution without assuming Gaussianity. Before turning to the discussion of phase transitions, we illustrate the main features of the CRLB and MCMC with a very simple example. We study a linear model that admits a straightforward analytical treatment, enabling a direct comparison of the two approaches.

\begin{figure}[t]
    \centering
    \includegraphics[width=0.45\textwidth]{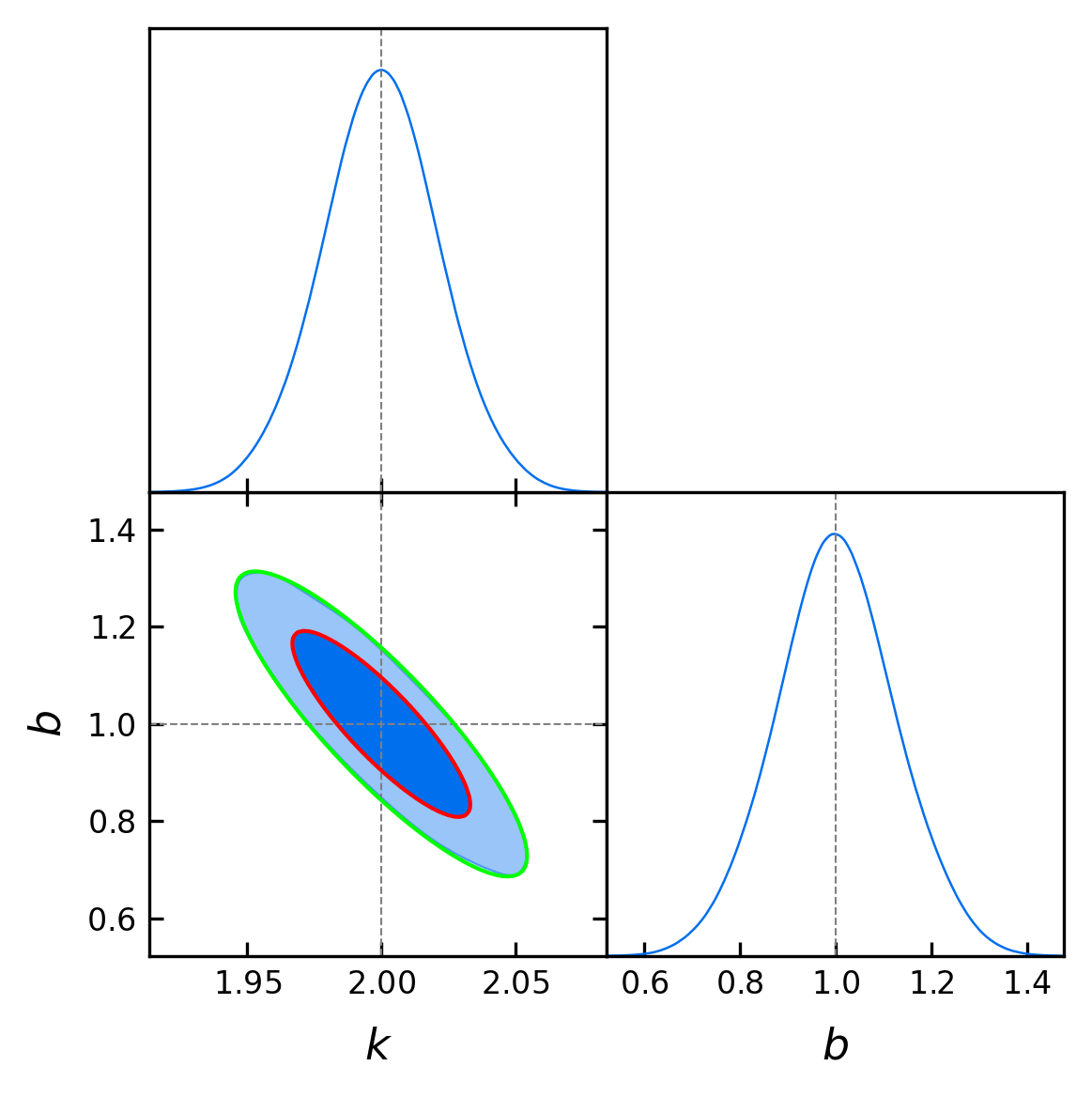}
    \includegraphics[width=0.45\textwidth]{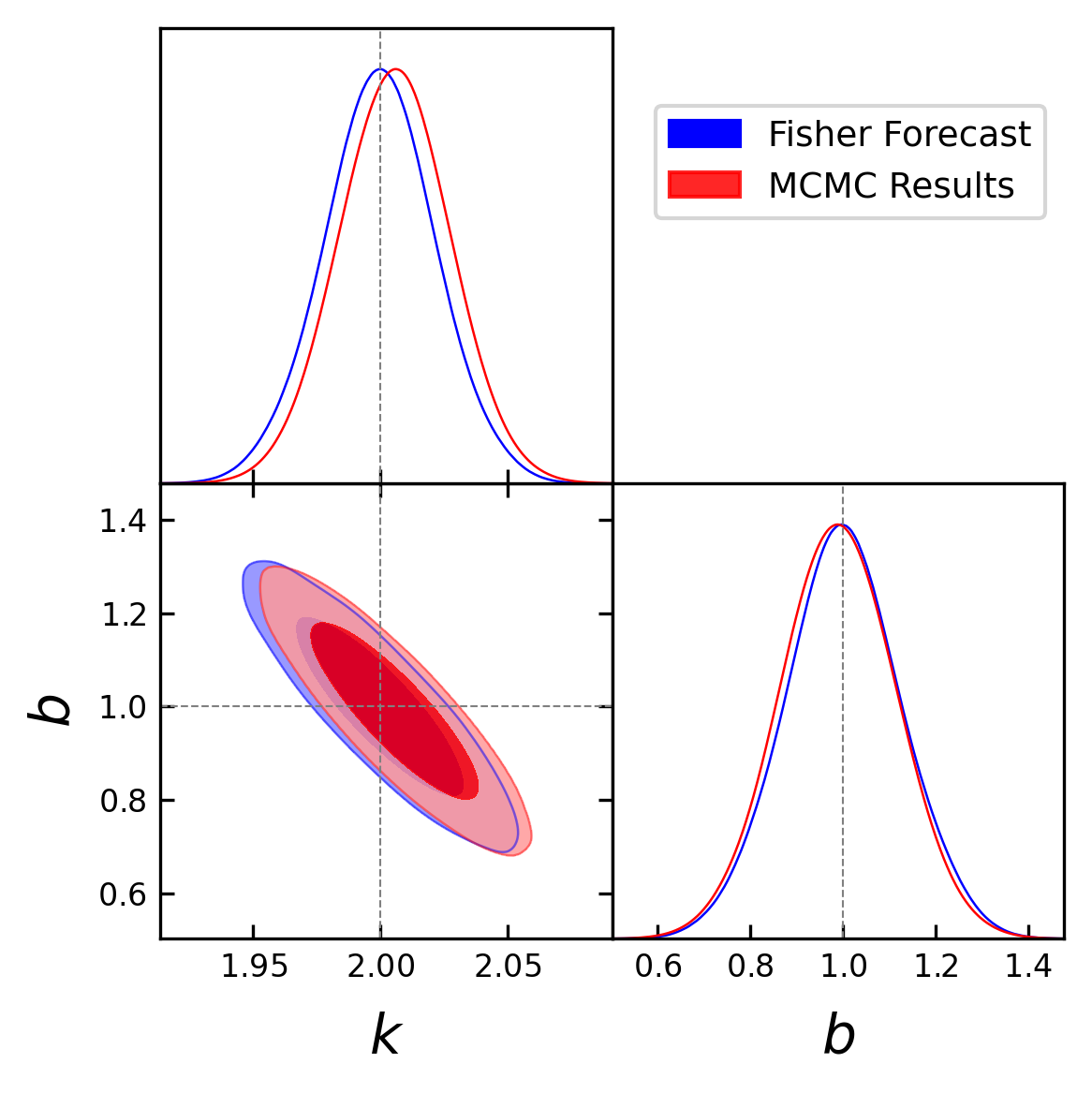}
    \caption{
    \label{poh}
    Comparison between the parameter constraints obtained from MCMC sampling and the confidence ellipses predicted by the FIM. The injected parameter values are $k = 2$ and $b = 1$, as indicated by the dashed lines. The darker and lighter shaded regions correspond to the 68\% and 95\% confidence intervals, respectively. In the left panel, the confidence ellipses (blue contours) are generated by drawing samples from a multivariate Gaussian distribution using the covariance matrix derived from the FIM. Superimposed red and green curves denote the analytical 68\% and 95\% Mahalanobis distance contours, respectively, and show excellent agreement with the sampled ellipses, thereby validating the correctness of the numerical Fisher-based approach. In the right panel, we present a direct comparison between the MCMC posterior distributions (in red) and the FIM forecasts (in blue). The overlap and slight deviations highlight both the validity and the limitations of the Gaussian assumption in the Fisher formalism. In this example, the minor discrepancy is likely caused by statistical fluctuations, while in more complex models with non-Gaussian posteriors, such deviations could become more significant.
    }
\end{figure}

In this linear model, the parameters $k$ and $b$ are inferred from simulated data. The mock data are assumed to follow a Gaussian distribution, and the corresponding log-likelihood function is given by
\begin{equation}\label{ugt}
    \ln \mathcal{L} 
    =
    -\frac{n}{2}\ln\left(2\pi\sigma^2\right)
    -\frac{1}{2\sigma^2}\sum_{i=1}^{n}(y_{d_i}-y_i)^2 ,
\end{equation}
where $n$ is the number of data points, $y_{d_i}$ denotes the simulated data, $y_i$ is given by $y_i = kx_i + b$ and $x_i$ are evenly spaced in the interval $[0,10]$. For simplicity, we fix the standard deviation to $\sigma = 2$. The uncertainties of the parameters are then estimated from the inverse FIM as
\begin{equation}
    \frac{\Delta k}{k}=\frac{\sqrt{\left(F^{-1}\right)_{kk}}}{k}
    ,\qquad
    \frac{\Delta b}{b}=\frac{\sqrt{\left(F^{-1}\right)_{bb}}}{b} ,
\end{equation}
where $k$ and $b$ are the input parameter values. The FIM can be computed analytically from Eq.~\eqref{ugt} as
\begin{equation}
    F=\frac{1}{\sigma^2}\sum\limits_{i=1}^n
    \left(
    \begin{matrix}
        x_i^2 & x_i \\
        x_i & 1
    \end{matrix}
    \right).
\end{equation}

The resulting relative uncertainties of the parameters $k$ and $b$, as derived from the inverse FIM, are presented in Fig.~\ref{fig:1} as functions of the total number of data points $n$. The figure illustrates how the precision of parameter estimation improves with increasing data size. This behavior reflects the intuitive expectation that larger datasets provide more information and thereby tighten the constraints on the model parameters. 

To further assess the validity of the FIM approximation, we compare the confidence regions predicted by the FIM with those obtained from full posterior sampling using the MCMC method. This comparison allows us to evaluate the accuracy of the Gaussian approximation inherent in the Fisher formalism. The resulting confidence contours for the parameters $k$ and $b$ are shown in Fig.~\ref{poh}.

The comparison shown in Fig.~\ref{poh} highlights the consistency between the FIM prediction and the posterior distribution obtained through MCMC sampling. For this linear model, where the likelihood is Gaussian and the parameter dependencies are linear, the posterior distribution is expected to follow a multivariate Gaussian form. As a result, the confidence ellipses derived from the inverse FIM (which approximate the CRLB) align closely with the contours of the MCMC-derived posterior. In addition, the injected parameter values, $k = 2$ and $b = 1$, lie near the center of both the 68\% and 95\% confidence regions, validating the reliability of the estimation procedure. The symmetry and orientation of the confidence ellipses indicate the degree of correlation between $k$ and $b$. This consistency also underscores the key assumption under which the FIM formalism is valid: the likelihood function is well approximated by a Gaussian near its maximum. In more complex nonlinear models, or under lower SNR conditions, deviations between FIM predictions and MCMC results become more pronounced, necessitating a full Bayesian treatment.

In summary, this example demonstrates that the FIM approach provides reliable forecasts for parameter uncertainties in idealized Gaussian scenarios and serves as a useful benchmark against which full Bayesian inference methods can be validated.

\bibliographystyle{JHEP}
\bibliography{refer}

\end{document}